\documentclass[aps,prl,twocolumn,superscriptaddress,groupedaddress]{revtex4}
\usepackage{subfig}
\usepackage{graphicx}  
\usepackage{dcolumn}   
\usepackage{bm}        
\usepackage{amssymb}   
\usepackage{slashed}
\usepackage{graphicx}				
\usepackage{amsmath}
\usepackage{mathtools}
\usepackage{tikz,pgf}
\usepackage{comment}
\usepackage{asymptote}
\usetikzlibrary{arrows,backgrounds}
\usetikzlibrary{fit,scopes,calc,matrix,positioning,decorations.pathmorphing}
\usepackage[all]{xy}
\usepackage{yfonts}

\newcommand{\Bracket}[1]{\ensuremath{\left\langle#1\right\rangle}}

\begin{document}
\title{(Pseudo-)Synthetic BRST quantisation of the bosonic string and the higher quantum origin of dualities}
\author{Andrei T. Patrascu}
\address{
email: andrei.patrascu.11@alumni.ucl.ac.uk
}
\begin{abstract}
In this article I am arguing in favour of the hypothesis that the origin of gauge and string dualities in general can be found in a higher-categorical interpretation of basic quantum mechanics. 
It is interesting to observe that the Galilei group has a non-trivial cohomology, while the Lorentz/Poincare group has trivial cohomology. When we constructed quantum mechanics, we noticed the non-trivial cohomology structure of the Galilei group and hence, we required for a proper quantisation procedure that would be compatible with the symmetry group of our theory, to go to a central extension of the Galilei group universal covering by co-cycle. This would be the Bargmann group. However, Nature didn't choose this path. Instead in nature, the Galilei group is not realised, while the Lorentz group is. The fact that the Galilei group has topological obstructions leads to a central charge, the mass, and a superselection rule, required to implement the Galilei symmetry, that forbids transitions between states of different mass. The topological structure of the Lorentz group however lacks such an obstruction, and hence allows for transitions between states of different mass. The connectivity structure of the Lorentz group as opposed to that of the Galilei group can be interpreted in the sense of an ER=EPR duality for the topological space associated to group cohomology. In string theory we started with the Witt algebra, and due to similar quantisation issues, we employed the central extension by co-cycle to obtain the Virasoro algebra. This is a unique extension for orientation preserving diffeomorphisms on a circle, but there is no reason to believe that, at the high energy domain in physics where this would apply, we do not have a totally different structure altogether and the degrees of freedom present there would require something vastly more general and global. 
\end{abstract}
\maketitle
\section{introduction}
String theory is sprinkled with a very complex web of dualities, including T-duality, S-duality, various target space dualities, Mirror symmetry [1], [2], [3], etc. Often theories defined in terms of very different geometrical and topological structures are found to describe "the same physics" at the level of the properties of the spectra of states emerging from such theories. The main argument is that many if not all of those dualities can be understood from the perspective of a higher quantum mechanics, in the same way in which gauge symmetries can be understood from the perspective of standard quantum mechanics. 
First we have to understand some foundational aspects related to quantum mechanics and its principles. There is a vast mathematical structure underpinning quantum mechanics, from the construction of a Hilbert space, to the commutation relation between observables and their promotion to the level of operators that are often matrix-valued and produce several possible outcomes instead of a single outcome as is the case in classical physics. The famous uncertainty implied by the construction of quantum mechanics seems to be a foundational aspect of nature. Indeed, non-commuting observables do not have commonly defined well determined values due to the nature of the commutation relations between them. As, axiomatically, the commutator relation is a derivative, obeying the usual Leibniz axioms of derivative 
\begin{equation}
[A,\cdot]=D_{A}
\end{equation}
it is worth keeping in mind that the non-commutative nature of the quantum observables can be described in terms of some form of parallel transport. 
I believe that the indeterminacies of quantum mechanics have a deeper cause. It is worth mentioning that such indeterminacies are still fundamental in nature, and cannot be removed by means of any hidden variable models one may consider. However, their ultimate cause lies in another aspect of quantum mechanics which I consider to be fundamental. 
\par In geometry there is always a tension between local and global structures [4]. We can consider a generic manifold to be locally flat, but the global structure will impose some restrictions on the functions that can be represented on it, restrictions that are not easily derivable in a strictly local manner. This tension between local and global structures has been a leitmotiv in the development of physics during the past centuries. The transition from the Galileo transformations and group to the Lorentz group brought in more structure not detectable at some small velocity, the transition further to curved spacetime led to the introduction of the global structure of curvature and its tensions with the local approximations of functions defined on the manifold, etc. 
\par Indeed, it is interesting to think of quantum mechanics as a way of harmonising two important principles of physics. First the Lorentz covariance and the causal structure introduced by it, as well as the locality of interactions, and on the other side, the intrinsic mathematical connection between the local structure and the global structure of a manifold. If physicists of past age had paid more attention at the optimisation principles they used in their calculations, starting with the Maupertuis's principle and culminating with the Euler-Lagrange equations and finally the Hamilton-Jacobi equations in classical mechanics they would have noticed that in the process of any optimisation principle, one has to point out the interference of non-local effects onto the so called classical optimal trajectory. The optimisation is by all standards a non-local phenomenon, as one admits the existence of a manifold of which the classical path is some optimum, defined by the rest of the manifold itself. One cannot have an optimum, without defining what is it that that optimum is for. Of course, if one considers only special relativity and its causal structure, the contribution of the rest of the manifold to the optimum must be fundamentally limited to causal connections, and hence much if not all the information of the global structure of the manifold is lost. But in that case, we would have a very hard time to understand why precisely a certain type of optimum exists, and not another. Now, if the outcomes of all observables were strictly determined, as classical mechanics demands, then the tension with the global-to-local influences we see in geometry would be impossible to account for in physics, and indeed, even the classical Hamilton-Jacobi equation would lose much of its predictive power. It is in a sense funny to notice that indeterminacy of observational outcomes is required for information coming from the global structure of the manifold to be able to determine the most likely optimal path obtained by classical Maupertuis type principles. We can notice that this is so by simply analysing an optimisation algorithm and observing that various directions must be continuously tried and discarded to obtain some transition towards an optimum. This implies the probing of the global structure of the manifold, or at least the structure of the manifold in the vicinity of the classical solution. If the optimal solution had been uniquely determined, probing the surrounding of it would bring us no closer to finding it, and such a surrounding would not even be present. Experience shows that optimisation principles are more general, offering solutions even in situations in which an exact focus on finding the specific solutions of the equations of motion is futile.
\par In any case, as a conceptual link between local and global structure exists, and is covered by a plethora of mathematical tools, nature must have a way of harmonising these connections. On one side, Lorentz invariance doesn't allow for signalling in a non-causal manner, re-enforcing some concept of locality, but on the other side, geometry implies "subtle" influences coming from the global structure of our manifold, which however must be in accord with Lorentz invariance and the principle of locality. Nature's "solution" to this problem was in a sense quite variate but had always a common denominator: global structure influenced local physics in ways that did not depend on Lorentz type signalling in a direct manner. The most acute aspect of this harmonisation appears in the black hole information paradox, but the situation is otherwise quite general. Instead of influencing the local structure directly, by means of causal effects, the influence came in the form of probabilistic effects and correlations. And here we come to the defining, in my opinion, property of quantum mechanics: quantum mechanics has the ability of encoding global information into local physical effects by means of probability amplitudes and quantum phases. In a sense it is quite obvious: the way in which quantum mechanics analyses all regions of the manifold that are physically accessible, even though they may be separated by energy barriers or light cones, does not contradict the notion of Lorentz causality, as it implies just correlations, and not causal signalling. Quantum mechanics, all by itself, gives us access to features that are causally separated or that are behind large energy barriers, simply by allowing for the global structure of our underlying manifold to influence the probability amplitudes (or maximal states of knowledge, according to Schrodinger) of the phenomena we try to determine locally. If such superpositions of states of knowledge were not possible, as would be the case in a classical approach to probabilistic calculations, we wouldn't gain the same access to global information as would be required from a purely geometrical point of view. That quantum mechanics can and indeed does give us insights into non-perturbative or topological effects should be well known by now. When writing Feynman diagrams in higher perturbative orders, we obtain so called "quantum" or "radiative" corrections, which basically imply some form of integration over inner momenta loops which exist due to the special relativistic effect of particle creation and annihilation. However, those particles, in the perturbative approach, are off-shell, hence they appear to have different masses than their physically realised counterparts. We have to distinguish the perturbative effects in this calculation and the quantum effects. The series expansion is usually performed in terms of a small coupling, while quantum effects behave, perturbatively, also similarly. However, the effects are different. The higher we go in the series expansion in terms of the coupling, the more quantum effects we take into account, and those quantum effects contribute to the probing, by means of integration over all momenta allowed for the inner loops, of the space of states at energies way beyond what we would otherwise access. In fact various particles have been and are expected to be detected by means of their off-shell amplitude contributions emerging in perturbative effects that would occur far below the energies where the on-shell counterparts would become real. While this should really be not surprising, the fact that we can link this with a quantum probing of the non-perturbative regions, by means of a perturbation theory that would otherwise look quite remote from the region where the inner loop particles are on-shell, signals an interesting connection quantum mechanics offers us to non-perturbative effects. Of course, the fact that quantum mechanics can probe into remote regions of our manifold can be studied also by means of the tunnel effect, or of the famous SU(2) anomaly for fermions, as identified by Witten. Therefore, it seems clear that the deeper reason for the indeterminacies of quantum mechanics is that the global structures of the underlying manifold must have an effect on its local structure together with its locally defined functions, an effect that is not due to a causal, light-signal-based connection, but instead is encoded in correlation effects that couldn't emerge unless quantum mechanics had its uncertainties. 
How can this help us understand dualities? In particular what does it mean to dualities in string theory? 
To explain this one has to go through various ways in which global information can be obtained locally. Our understanding of the physical phenomena at lower energies is based on the assumption that the fundamental object we must deal with is a particle. A particle is a zero-dimensional object, and as such, it has no dimension whatsoever. This concept has been challenged in various ways and for good reason. 

\par One way in which it has been challenged, which will be of importance later on, is in topology. There, a construction called (co)homology has been introduced as a topological invariant, namely a mathematical object that allows us to classify objects that were equivalent from the point of view of smooth transformations that did not take into account the notion of distance (or metric). The invariant had to be able to get the same value anywhere on the objects that were equivalent in terms of those smooth transformations, and discretely change in value, in a predictable way, on objects that could not be deformed into each other by smooth transformations. The result in this case was the classification of tori in terms of the number of handles or the number of holes. In any case, such invariants also have some form of indeterminacy. That very indeterminacy is the same we see in quantum mechanics, and the reasons are the same. It was observed that in order to define a simplicial complex that would triangulate a space, we have to form linear combinations of certain simplexes, and that those linear combinations had well defined coefficients. The (co)homology, and the rest of the invariants were by construction determined only up to the coefficient structure, which is usually denoted as the coefficient structure of (co)homology. Such a coefficient structure is determined by one of the axioms of (co)homology, namely the so called "dimension axiom" which fixes the fact that the (co)homology of a point space is trivial. Topologists call "trivial" the cohomology structure that is everywhere reducible leading to it being $\mathbb{Z}$ in the order of the dimension of the point, namely at order $n=0$ and zero everywhere else, hence all "regions" of a single point space are connected among themselves and contractible to a point (which seems trivially true). Well, that would definitely be a point in our standard understanding. It was however soon realised by some mathematicians that a specific choice of a coefficient structure limits the ability of a (co)homology to detect certain features, and that one cannot, by definition, construct a (co)homology without a coefficient structure. The remaining alternative was to abandon the dimension axiom and introduce additional structure into the point. If we introduce non-trivial coefficients instead of simply $\mathbb{Z}$ in order zero of the (co)homology of a point, it means we change the coefficients structure of the simplicial complex from which we derived the (co)homology. This amounts to a cohomology that would have structure attached to the points, and if even higher orders in cohomology would receive structure, instead of being simply zero, more information could be attached to our point. However, no matter how much information we attach to a point, there will always be properties of the actual topological space that a certain invariant won't be able to detect, at the expense of others that it will be able to detect. 
\par This should remind us to some extent to quantum mechanics. We will notice that if we extend the point to a more complex object, we will do it at the expense of detecting certain features that can be detected by points, but will get to the situation in which we will be able to detect structures not detectable by points. String theory does precisely this. 
\par First, strings are extended objects, therefore they will have some form of access to non-local information, more than a non-extended fundamental object. That means we will obtain some access to non-perturbative information by means of the dimensional extension we applied to even a classical string. In that sense, even a classical string starts by having some insight into quantum phenomena. 
\par Dualities are instances in which string theories defined in radically different geometric and topological contexts provide the same physical states and symmetries between them. The reason why we observe this is because of a subtle interplay between local and global information as detected by a string instead of a point. Indeed, the information obtained by a string in a certain topological and geometrical context may as well be totally identifiable with that obtained in another context. Is there a relation between such dualities? Or, better stated, would there be possible to determine new dualities by simply analysing some properties of quantum mechanics? It seems this would be the case. 
\par As mentioned previously, quantum mechanics, and in particular its fundamental indeterminacies, are the way in which nature harmonises the relation between global shapes of our manifolds and the local nature of interactions. In the context of extended objects, a similar effect must take place. 
Quantum mechanics introduces the additional quantum phase. Differences in local phases of various components of the probability amplitude give us probabilistic insight into the global structure of our manifold. Basically, in quantum mechanics we are able to probe the non-perturbative structure by means of relative differences in the same complex phase. Entanglement is the result of the existence of quantum states with complex phases that amount to spacial (or otherwise) inseparable descriptions of the phenomena. If we were able to expand such entanglement to extended objects, we should in principle find that certain string theory configurations involving various global and local structures become either equivalent or strictly dependent on each other. This is the origin of string theoretical dualities. 
These are the main claims of this article. Let us see in what follows what can be proved and to what extent. 
\section{target dualities}
The fact that quantum uncertainty and quantum statistical fluctuations allow for non-local information to be revealed is relatively obvious. The fact that inner (say loop particle, in Feynman diagrams) states are not practically realised but their possible existence, even at remote regions in spacetime or vastly separated in energy from the experimental region, do affect the probability amplitudes, makes it relatively clear that global information finds its way in our quantum description. Moreover, as seen by means of the fermion SU(2) anomaly, quantum mechanics can also probe topologically non-trivial regions of symmetry groups, allowing us to figure out what quantum field theories are meaningful and realised in nature. 
\par The obvious question one may ask is: would it be possible to use quantum mechanical arguments to determine what topological structure a symmetry group should have? Isn't it possible to use quantum arguments to find arguments for changing our understanding of the fundamental symmetry groups of nature, for example like in the transition from Galilei groups to Lorentz groups introduced by Einstein? 
\par Let us see what we can say about how global information is recovered locally in another framework, that of conformal field theories, in particular conformal field theories defined on a certain 2-dimensional manifold $\Sigma$ which may be embedded into a target manifold we will discuss about soon. The Lagrangian $L$ of such a theory is fully solvable, particularly due to the infinite symmetry group that exists in two dimensions. That means we can determine the Virasoro central charge $c$ of the theory, giving us the quantum conformal anomaly, the allowed states and their corresponding operators, as well as their operator product expansion coefficients. This Lagrangian is defined by means of its couplings and the variation of the couplings due to renormalisation group flows show us how the operators associated to those couplings behave in various limits. We can deform this Lagrangian by adding to it more couplings and operator pairs. 
\begin{equation}
L\rightarrow L'+\Bracket{g_{i}|\mathcal{O}}
\end{equation}
where we can simply express 
\begin{equation}
\Bracket{g_{i}|\mathcal{O}}=\sum_{i}g_{i}\cdot f_{i}(z,\bar{z})
\end{equation}
If the modified Lagrangian also corresponds to a conformal field theory then we found a potential family of conformal field theories. According to the dimension of the operators, we have irrelevant, marginal, or relevant operators, corresponding to how their couplings flow in the IR. The marginal operators have the operator dimension exactly $2$ and hence preserve the classical scale invariance and the couplings remain dimensionless. However, for marginal operators the couplings may also change under renormalisation flows. If the addition of such marginal operators preserves the dimension of the couplings, then we call those operators truly marginal and we obtain a spectrum of Lagrangians, and hence a "space" of theories. On this space we can imagine a group with some symmetry operations acting on the space of Lagrangians, modifying for example $L_{1}\rightarrow L_{2}$ and moving from point to point on this space of theories. We can imagine various ways of classifying and partitioning this space, one of them being by a subgroup of our group of transformations, call it $\mathcal{G}_{D}\subset \mathcal{G}$ which transforms our theory into an equivalent theory. The question of what means "equivalent" here is quite interesting. The theories should in principle be physically equivalent, and hence should provide the same physically observable answers for similar contexts. But such a space of theories, let us call it $\mathcal{M}$ is clearly not detached from the structures one can define on it. It would be therefore imaginable that an equivalent of a wavefunction could be defined on such a space and that, given certain types of incompatibility relations between theories, analogue to the non-commutativity relations of quantum observables, we would obtain a form of higher entanglement that would play a similar role for describing distinct but inseparable theories. As the wavefunction and its inherent uncertainty allows us to statistically probe the non-local, non-perturbative regions of the usual phase space, in this case, we would be able to find inseparable and hence strongly correlated theories that would amount basically to what we are accustomed to call "dualities". In this context, dualities would be nothing but strongly entangled "states" in this space of theories. This aspect can be made decently precise if we think about the connection between gauge symmetry and quantum mechanics and how they both have the same origin. Adding more theories in the conformal field theoretical construction amounts to adding more operators and couplings. But the overall context is that of a string theory of which the conformal field theory is the worldsheet representation and for which the background in which it moves is represented by yet another space that is (at least to a high degree) determined by the properties of the worldsheet. It is well known that in this context, the couplings $g_{i}$ on the conformal side correspond to allowed target space backgrounds in which the string can propagate. The couplings in the conformal worldsheet can be re-organised into the usual gauge fields of the background theory. This makes the conformal coupling flow equations basically the equations of motion of the higher gauge fields in the background description. If we think of a simple bosonic string theory, the couplings are all organised into three fundamental gauge fields: the background metric, let us call it $\hat{G}_{ij}(X)$, the anti-symmetric tensor field $\hat{B}_{ij}(X)$ and the dilaton field $\hat{\Phi}(X)$ where $X$ is the background or target space coordinate.
Then we can write the worldsheet action as 
\begin{widetext}
\begin{equation}
S=\frac{1}{4\pi\alpha'}\int_{0}^{2\pi}d\sigma\int d\tau [\sqrt{g} g^{\alpha\beta}\hat{G}_{ij}(X)\partial_{\alpha}X^{i}\partial_{\beta}X^{j}+\epsilon^{\alpha\beta}\hat{B}{ij}(X)\partial_{\alpha}X^{i}\partial_{\beta}X^{j}-\frac{\alpha'}{2}\sqrt{g}\hat{\Phi}(X)R^{(2)}]
\end{equation} 
\end{widetext}
No wonder our theories in the conformal space $\mathcal{M}$ are entangled, or otherwise strongly correlated, given that the space itself can be reorganised into a set of just three distinct gauge fields. We have used the standard notation for the worldsheet metric $g_{\alpha\beta}$ and the associated determinant $g=det(g_{\alpha\beta})$ while $\alpha'$ is the string coupling constant proportional to the inverse of the string length and $R^{(2)}$ is the worldsheet scalar curvature. 
\par We already identified the uncertainty of this construction. In certain cases, the addition of different $\Bracket{g_{i}|\mathcal{O}}$ terms does not make the theory properly distinguishable from another. Moreover, the various combinations of terms that can enter in that pairing may vary from one addition to another, and hence the sum introduced above will be over different pairs of couplings and operators. This looks like a gauge/quantum indeterminacy. For all practical purposes, when we perform a Feynman path integral we integrate over all potentially equivalent trajectories that lead to the same final state. The trajectories may look different, but from the point of view of the experimental setup they are not distinguishable. In order to distinguish them we need to modify the experiment in order to add some "which path" detector, leading to a totally different context. We can imagine the same thing here. In a sense, string theory in the target/background space is the final "interference pattern" outcome of our Feynman style integration over all possible paths in the theory space. The uncertainty in choosing a path amounts to the types of string theory couplings and in the type of target space configurations we can construct. In a sense, this is the way in which the target spacetime "emerges" from "entanglement".
\par We are accustomed to see dualities as symmetries occurring in such a construction, namely symmetry operations from our group $\mathcal{G}_{D}$ that lead us to equivalent theories. It seems that there is much more structure to this than just being a group. If we look at the string theoretical description, then the worldsheet action allows us to calculate $S$-matrix elements for scattering in the target space. The objects that will scatter will be fluctuations in the target theory around the classical values of the background fields $\hat{G}_{ij}$, $\hat{B}_{ij}$ and $\hat{\Phi}$. We may denote them for convenience as 
\begin{equation}
\begin{array}{c}
\tilde{G}_{ij}(X)=\hat{G}_{ij}(X)+h_{ij}(X)\\
\\
\tilde{B}_{ij}(X)=\hat{B}_{ij}(X)+b_{ij}(X)\\
\\
\tilde{\Phi}_{ij}(X)=\hat{\Phi}_{ij}(X)+\phi_{ij}(X)\\
\end{array}
\end{equation}
where the fluctuations $h_{ij}$, $b_{ij}$, $\phi$ describe the massless graviton, antisymmetric tensor and dilaton of the target space theory. Truly marginal operators on the worldsheet correspond to massless particles in the target space. Relevant worldsheet operators correspond to instabilities in the target space and hence the existence of tachyons in the background, while irrelevant operators on the conformal side correspond to massive target states. In string theory those operators are always dressed in such a way that they appear as vertex operators with dimension $(1,1)$ (denoting the left- and right- handed conformal dimension of the operator). In general we say there exists a toroidal compactification in the background if for some reason $d$ of the total background dimensions have been compactified and the gauge fields $\hat{G}$, $\hat{B}$, $\hat{\Phi}$ as well as $X$ are independent. The general assumption is that the operators in $\mathcal{G}_{D}$ correspond to target space dualities. 

\par In standard quantum mechanics we can describe the fact that observables are incompatible by means of commutation relations. Let us look at this situation in a broader sense now. After we promoted observables from mathematical structures having a unique possible outcome, that is in a one-to-one relation to the "real" property of our system, to operator valued observables in which more possible outcomes exist, that are obtained as possible answers to questions regarding the property of the system as described by the observable, we realise that not all such "multi-valued" observables can be simultaneously determined to arbitrary precision. That means, if we alter the experimental setup and hence we make the question we ask precise enough to determine one of the properties, another property may become completely undetermined in this new setup. This is essential for the formulation of quantum mechanics, and it is most generally expressed in terms of the possibility of accurately determining the answers of two different questions by the same experimental setup. Mathematically this amounts to a splitting of the phase space into mutually non-commuting variables (say position and momentum) and the implementation of a non-trivial generalised derivative as being non-trivial, namely $D_{A}=[A,\cdot]$. But as we can see now, there are various mathematical ways in which such an incompatibility of observables can be expressed, and each applies best to a specific domain. What would be the best way of applying it to the space of theories? 

\par The existence of incompatible observables is the starting point for the so called "quantum fluctuations" of the outcomes of some of the observables, in any physical context. But it is precisely this type of fluctuation that allows us to have access to global data in the form of correlations that would otherwise not be possible. We cannot directly probe the global structure of the manifold we are working on, but we can probe it by means of correlations that amount to the encoding of the global structure in fluctuations of the outcomes. We do something very similar in the context of string theory, where the quantisation of a string follows a very similar pattern, with some distinctions related to the types of algebras emerging. In effect, we start with a Witt algebra in a classical context, describing the meromorphic vector fields on a Riemann sphere, or the complexification of the Lie algebra of polynomial vector fields on a circle, and by quantisation we obtain the deformed Virasoro algebra which is the unique central extension of the Witt algebra which includes the central charge of the conformal field theory, namely the so called "quantum conformal anomaly". This deviation from the Witt algebra in the form of an extension makes sense because quantum mechanics is sensitive to global structures that are now properly encoded in the quantisation algebra. But is it certain that this extended algebra represents the right global structure? The original idea related to extensions was defined by an early seemingly simple question: given two groups $G$ and $H$, what would be the groups that we can form out of those two groups, say $E$, that would have their normal subgroup $N$ isomorphic to $G$ and the quotient group $E/N$ isomorphic to $H$? A normal subgroup is a subgroup that is invariant under the application of all elements of the larger group in which it is a normal subgroup, in this case, all elements of $E$ would leave $N$ invariant, but the classification of $E$ in terms of elements of $N$ (by means of the quotient operation) would amount to the group $H$. If we think in terms of global effects in quantum mechanics, it is interesting to note that in general in physics, the symmetry groups of quantised systems are central extensions of their classical counterparts. Such central extensions are constructed "by co-cycle" which basically means that they do indeed refer to the global, cohomology-related properties of the groups involved. Moreover, central charges, as the ones we detect in the quantisation for conformal field theories, arise naturally due to the requirement of forming a Lie algebra using projective group representations to start with. Consider we have a group structure, and we wish to characterise it. We can for example translate it into a projective representation. For a group $G$ acting on a vector space $V$ and given an underlying field $F$, we can find a group homomorphism 
\begin{equation}
G\rightarrow PGL(V)=GL(V)/F^{*}
\end{equation}
where $F^{*}$ is the normal subgroup of non-zero scalar multiples of the identity. We would have a projective representation of a group $G$ given by the operators $\rho(g)\in GL(V)$, $g\in G$ where the homomorphism property is obeyed up to a constant
\begin{equation}
\rho(g)\rho(h)=c(g,h)\rho(gh)
\end{equation}
and our "phase" $c(g,h)\in F$ is in fact a co-cycle. To understand the topological nature of this statement, one can find a set of linear operators $\tilde{\rho}(g)\in PGL(V)$, $g\in G$ with the property that
\begin{equation}
\tilde{\rho}(gh)=\tilde{\rho}(g)\tilde{\rho}(h)
\end{equation}
The two situations are different. On one side the homomorphism relation must ignore the phase, on the other, we discuss about equivalence classes in which the phase is merged into the definition of the representation. Therefore $\tilde{\rho}$ forms a class of operators, and there will be topological obstructions emerging in the following way. We can of course pick from each such equivalence class a representative $\rho(g)\in \tilde{\rho}(g)$ such that, at the level of those representatives, the homomorphism rule is obeyed exactly, and not only up to a phase. This is called a de-projectivized operator that obeys 
\begin{equation}
\rho(g)\cdot \rho(h)=\rho(gh)
\end{equation}
In general when we have a projective representation of a group, it is not always possible to go back to a linear representation meaning we cannot go from a map $\rho: G\rightarrow PGL(V)$ to a map $\rho: G\rightarrow GL(V)$, the obstruction to this transition being characterised by group cohomology. Quantum mechanics usually takes into account the phase, which appears as a global effect in the groups, and therefore, in order to quantise theories in general, and in particular conformal field theories, what we do is to find another, related group that can be lifted towards a general linear group. This different group is in fact the central extension of the original group, which is a group $H$ appearing as a subgroup of $G\times GL(V)$ defined as 
\begin{equation}
H=\{(g,A)\in G\times GL(V)|\pi(A)=\rho(g)\}
\end{equation}
$\pi: GL(V)\rightarrow PGL(V)$ being the quotient map from $GL(V)$ onto $PGL(V)$. $H$ is a subgroup of $G\times GL(V)$. 
The connection between projective representations of symmetry groups and their de-projectivisation is quite interesting as it involved topological group structure which has to be taken into account. By Wigner's theorem for example, if $G$ is a symmetry group, it can be represented projectively on a Hilbert space by unitary (or anti-unitary) operators. This usually implies transitioning to the universal cover group of $G$ and taking it as a symmetry group. As we have noticed, this works for the Lorentz group but does not quite work for the Galilei group. Instead, to establish the symmetry group of the Schrodinger equation one has to construct the central extension of the Galilei group, known as the Bargmann group. The reason for this is a topological obstruction in the very construction of the Galilei group, in particular a non-trivial group cohomology feature that makes the transition to quantum mechanics not as direct as in the case of the Lorentz group. In a sense, if we first learned of quantum mechanics, we could have thought to re-consider the reference frame invariance transformations in order to avoid the topological obstruction that emerged in the Galilei group. We could have therefore constructed the Lorentz group by simply demanding that the group cohomology be trivial in order to make it compatible with our quantum formulation. It so happened however, that we first discovered the Galilei group, and only thereafter quantum mechanics. But it also so happened that while we decided to trade the Galilei group for the Lorentz group when passing from a classical to a quantum construction (despite the fact that we still work with Schrodinger's equation, we know it is only a low energy approximation), in the case of string theory we sticked to the Witt algebra and decided to construct its extension by a 2-cocycle to obtain the Virasoro algebra. This is of course technically fine, but then, what would have happened if we discovered first quantum mechanics and then we tried to define what the proper string theory would be? Would we still look for an extension of the Witt algebra? 
In a sense it seems natural (as in, obstruction free) to have the Lorentz group as a symmetry group if quantum mechanics is to be assumed, and that means quantum mechanics needs to take into account by its complex phase, aspects that the global Lorentz group determines. By going to a cohomology trivial group, we can construct quantum mechanics and allow for the de-projectisation and for the meaningful construction of the complex phase. However, thinking historically, we first came up with the Galilei group, which has a homological obstruction, and when we discovered quantum mechanics, we had to construct its extension, going to the Bargmann group in order to obtain Schrodinger's equation together with its Galilei invariance. We didn't keep the co-cycle extension of the Galilei group though while advancing in our understanding of nature. Instead, we replaced the Galilei group with the Lorentz and Poincare groups which are not presenting a cohomology obstruction (or a non-trivial cohomology). 
However, in string theory we start with the Witt algebra in the classical context, and in order to quantise, we perform a co-cycle extension of the Witt algebra to obtain the Virasoro algebra. Mathematically this seems a consistent way of working, but so did the Galilei group extension seem when looked at from the perspective of Schrodinger's equation. It's just so that Nature didn't do it that way. 
It would be pedagogical to remember what the situation would look like in the transition from Poincare/Lorentz group to the Galilei group (by the Wigner-Inonu contraction).
Having the generators $K_{i}=M_{0i}$ with $i=1,2,3$, we obtain 
\begin{equation}
[K_{i},P_{j}]=i(\eta_{0j}P_{i}-\eta_{ij}P_{0})=-i\delta_{ij}P_{0}
\end{equation}
and by the Wigner-Inonu contraction we obtain 
\begin{equation}
[K_{i},P_{j}]=-i\delta_{ij}M
\end{equation}
as in the non-relativistic limit the mass term will become dominant. 
\par We can look at the mass $M$ as an element of a Lie algebra that commutes with all other generators, also because it is a scalar, and hence cannot be removed by some continuous re-definition of the generators. It therefore amounts to a central extension of our Galilei group. In each irreducible representation of the Galilei group we therefore have to have this mass term, represented as a scalar, and it is not possible to connect two representations with different masses by unitary transformations. In terms of quantum mechanics, the wavefunction obtains always a phase factor corresponding to the mass, and the condition above imposes a superselection rule that implies that we cannot construct wavefunctions as linear superpositions of wavefunctions of particles with different masses. This must be so because the global phase is not observable in quantum mechanics, but the relative phase is. Now, boosts even in the Galilei group, do induce phase shifts. However, due to the fact that the global phase is not detectable in quantum mechanics, this does not affect the preservation of the Galilei group. However, if different masses were present in a linear combination each component would acquire a different phase, and a differential phase is detectable in quantum mechanics. To preserve the Galilei group structure, this cannot happen, and therefore different masses cannot appear in the same linear combination of states. This is an interesting argument, because such transitions between states of different masses are indeed observed in Nature, and hence a simple central extension of the Galilei group, as constructed by means of its mass as a "central charge" is not at all sufficient to describe Nature. The reason why we didn't allow the transitions between states of different masses in quantum mechanics is because quantum mechanics was in principle able to detect the topological structure of the group, namely the non-trivial cohomology associated to the Galilei group and the special type of central extension needed to construct the quantum wavefunction made that structure only more manifest. In fact, the obstruction to de-projectization of the projective reconstruction of the Galilei group amounted to quantum mechanics making it clear that there is a discrepancy between the groups we intended to use, and the real groups Nature chose. Interestingly enough, the Lorentz group, while topologically more trivial than the Galilei group, differs from the Galilei group by the way it behaves in the far reaches of its parameters, in particular in the far reaches of the velocity. That is the global structure that quantum mechanics detected, and there was also the reason why it presented an obstruction to a clear-cut construction of a quantum representation. 
\par In fact, within the Lorentz local group we can have various linear combinations and superpositions between particles of different masses, and we indeed can also entangle various of their properties. We could call such an effect a "duality" in the sense that the description of theories with different central charges may be related and the properties of the two theories may be "entangled" and non-trivially correlated. Would it be possible to search for a more generalised symmetry group that would be topologically trivial as opposed to one that would produce central charges (anomalies)? We have to ask then what can be done in order to create a structure that is indeed changing its topology?
\par Topology changes occur in string theory among dualities like T-duality with non-trivial H-fluxes, and that indicates that topological structure is not pre-defined nor essential for a theory to be consistent. In essence, in the case of the transition between Galilei and the central extension of the Galilei group, what we had to take into account was in particular a non-trivial cohomology of the group, which was avoided not by means of a central extension (as that would give us still a non-relativistic theory), but by means of the change of the original group itself, in particular, by considering it only a local approximation of another group valid at a different parameter range. In any case, Einstein was guided towards his construction of special relativity and towards the use of Lorentz transformations by clear experimental evidence, which we don't quite have in the case I wish to discuss now. There is however another aspect about cohomology groups that may play a role. As mentioned above, the cohomology groups are defined only up to their coefficient structures. The coefficient structure is the topological structure of a point space. We define a cohomology theory by specifying how it would behave for a single point space to lowest order. In the simplest case, it is usually given that 
\begin{equation}
\begin{array}{c}
H_{0}(P)=\mathbb{Z}\\
H_{n}(P)=0,\;\; \forall n>0\\
\end{array}
\end{equation}
This appears as the "dimension axiom" in the Eilenberg-Steenrood axioms of cohomology, and generalised cohomologies do in fact ignore this axiom. 
\par When ignoring it, additional structure is given to the point space in the cohomology theory, and the transition between cohomology theories with different coefficients is permitted by the Universal coefficient theorem, which is represented by an exact sequence depending at one point on the Extension groups, but not being uniquely defined by them. We can indeed trivialise the Extension by picking a specific coefficient structure. Maybe in order to determine a better foundation for the dualities in string theory, it is this route that we could take. Of course we can only guess what type of coefficient structure our point should have, but probably there is no pre-defined or pre-given coefficient structure, it is just that in some coefficient structures it is simply much harder to operate and to make certain processes or dualities manifest. However, due to the fact that universal coefficient theorems for generalised (co)homologies exist only in very particular cases, as basic relations, and not as truly universal rules, practical advances in this direction are still very limited and localised. 
\par I suggested that we could search for relations between different coefficients in cohomology as a source for dualities some time ago [5] but at that time the connection with quantum mechanics was not yet clear to me. In fact, it should have been obvious. The anomalies in quantum theories do usually emerge due to the fact that our symmetries are sometimes imprecise and quantum mechanics tends to probe and detect global structures of the groups we "offer" it to work with, making it quite manifest when the groups we present to it are not really what Nature intended. Usually we determine such issues early on as inconsistencies in the quantum descriptions or in the quantisation procedures that do appear in the form of anomalies, but it is never quite so easy to see how to extend or expand our groups or structures, and that not because of quantum mechanics, but because of our limited access to experimental evidence, and of course because we don't quite ask the questions for which such modifications of the groups would be pertinent. I suspect the hierarchy problem and the problem of the cosmological constant are basically questions that would be very pertinent to this.

\par In particular it is interesting to notice the existence of a quantum mechanical structure, of the form of superposition of probability amplitudes and of entanglement, as in non-separability due to the existence of such phase dependent probability amplitudes, at the level of the theories that can be connected by means of dualities. It seems to be possible to re-interpret even various mathematical structures by means of quantum type correlations, uncertainty, wavefunctions, etc. Dual theories could be for example re-interpreted in the form of "entangled" theories, namely theories that are not separable in a cartesian sense in their respective domain of parameters, given some additional higher algebraic structure that connects them and makes them not separable. In this sense, quantum mechanics may lie at the heart of dualities in gauge or string theories. In fact, if we look at the Galilei group and its extension, we notice the type of change that occurs if we set $M=0$. The central extension indeed becomes trivial, amounting to be just the extension by the direct product, a fully separable structure in itself and hence
\begin{equation}
M=0 \Rightarrow c(g',g)=1 \Rightarrow U(g')U(g)=U(g'\cdot g)
\end{equation}

Of course the construction of the extended Galilei group with $M\neq 0$ doesn't just yet allow for non-trivial matrix elements between different masses, due to the superselection rule, but that simply happens because the Galilei group ignores the global structure of our natural group, or, in a metaphorical sense, following ER=EPR, it doesn't take into account the "wormhole" connection between them given by the global structure of the Lorentz group. 

In the context of entanglement entropy and the ER=EPR duality, the problem of defining operator algebras has been discussed in [15] where bulk spacetime connectivity or lack thereof is associated to the operator algebraic structure of quantum gravity. This results in a construction that defines the structure of entanglement in a non-trivial situation. In [16] T-duality is used to connect a pair of black holes such that a wormhole forms where the throat is proportional to the zero-point (Planck) length. This is presented as an attempt to provide a geometric representation of the entanglement between particle-anti-particle pairs.
\par If we think in terms of possible extensions, as the direct product extension for $M=0$ indeed is trivial, it clearly appears that the use of the term "non-separability" or "entanglement" when discussing about such structures is not merely a matter of semantics, and indeed it has a deeper meaning. 
One of the questions that has rightfully asked itself is, if ER=EPR is valid [6], what is the low energy interpretation of such a phenomenon? We can imagine black holes being entangled and forming wormhole geometries between them, as Susskind showed in his toy model, but surely, given that we have indeed entanglement between particles that are not even close to being black holes, what is the "wormhole geometry" in that case? If we take as an initial model the Galilei group extension for $M=0$ becoming trivial or a direct product, then we can imagine that for $M\neq 0$ the obstruction of the Galilei group creates a situation in which no "wormhole" is present, and states of different masses are not connected and cannot appear in linear combinations of states. However, in the case of the Lorentz group, the cohomology is trivial, and hence a "connection" is created, leading to transitions between masses. In this sense, this ER=EPR relation can be regarded as particles being created or decaying. The "wormholes" in the low energy domain and in the context of group cohomology are nothing but the particle creation and annihilation processes permitted by relativistic quantum theory. 
\begin{figure}
  \includegraphics[width=\linewidth]{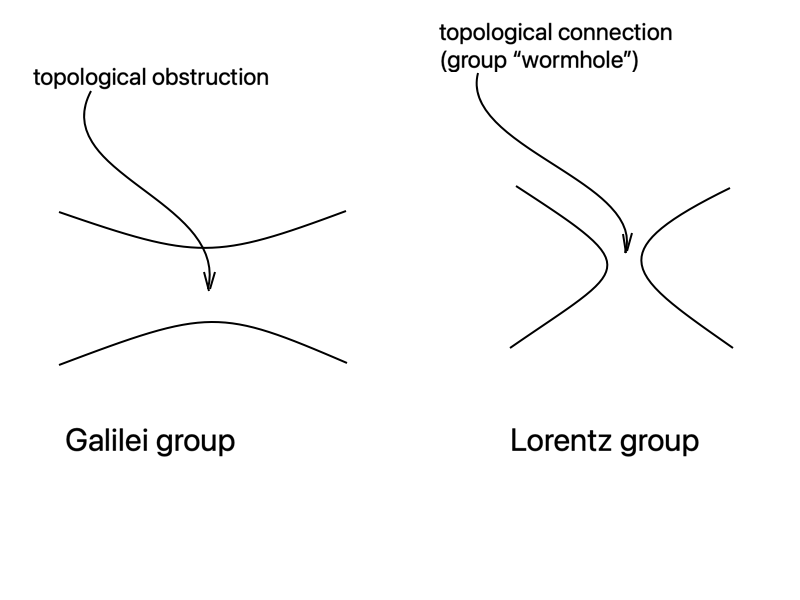}
  \caption{Topological obstruction of the Galilei group vs. topological connection of the Lorentz group}
  \label{fig:groups}
\end{figure}

The wormhole geometry is a geometrical representation of the fact that particles shared at some point a causal connection and their wavefunctions become superposed and eventually entangled, their description having to become global, the full information about one particle being encoded in the whole system containing the two particles. This can be translated in terms of cohomology with variable coefficients. In one representation we keep the particles as such and describe our theory with a regular cohomology theory with trivial coefficients, say $\mathbb{Z}$. This works as long as we take into account the proper extensions. After all, entangled particles do exist also in the Galilei group and can be described by Schrodinger's equation. If we do not wish to use an elaborate extension of our group, or that extension simply doesn't quite suite our requirements, for example due to inter-mass transitions, we may decide to change the extension, but by doing so we also have to change the coefficient structure accordingly. In fact, in some cases we can change the coefficient structure into a structure that involves for example elliptic curves, in which a global geometric connection between the particles would be more "manifest". This would of course be just a representation of an inner structure of a "point" transforming a system made out of two apparently distinct subsystems, bound however by a more complicated extension of their group, into one sole system, but with an intricate geometrical connection between them, manifest in the non-trivial point structure, but with a simpler group. 
In any case, the mathematics of universal coefficient theorems for generalised cohomologies is much more complicated and in many very important cases, the problems associated to it are not yet resolved. We can however apply the universal coefficient theorem in the case of ordinary cohomology and therefore we can show how ordinary (co)homology determined (co)homology with arbitrary coefficients. Given a chain complex $C_{*}$ of abelian groups and a field $F$ the homology group $H_{p}(C,F)$ and the cohomology group $H^{p}(C,F)$ are related by a dualisation $H^{p}(C,F)\sim Hom_{F}(H_{p}(C,F),F)$. If the coefficients do not form a field, but are an arbitrary abelian group, then this relation is corrected by an $Ext$ group. If we think in a dual sense, then if $F$ is a field then there is an isomorphism $H_{n}(C)\otimes F \sim H_{n}(C\otimes F)$ and if $F$ is more general, then we obtain corrections by the $Tor$ group. I will now give the standard construction of the Universal Coefficient Theorem [7]. 
If $C_{*}$ is a chain complex of free abelian groups and $A$ is an abelian group then we denote $C^{*}=Hom_{A}(C,A)$ the dual cochain complex with respect to $A$, $H_{n}(C)$ the chain homology of $C$ and $H^{n}(C,A)$ the cochain cohomology of $C^{*}$ with coefficients in $A$. There exists a canonical morphism of abelian groups 
\begin{equation}
\int_{-} (-): H^{n}(C,A)\rightarrow Hom_{A}(H_{n}(C, \mathbb{Z}),A)
\end{equation}
given by taking the map of a cocycle into the evaluation of that cocycle on a chain 
\begin{equation}
[\omega]\rightarrow ([\sigma]\rightarrow \int_{\sigma} \omega = \omega(\sigma))
\end{equation}
Then the universal coefficient theorem then states that there is a short exact sequence of the form 
\begin{widetext}
\begin{equation}
0\rightarrow Ext^{1}(H_{n-1}(C),A)\rightarrow H^{n}(C,A)\rightarrow Hom_{A}(H_{n}(C),A)\rightarrow 0
\end{equation}
\end{widetext}
and the sequence splits (albeit non-canonically). 
For group cohomology some additional information is needed. Suppose $G$ is a group and $A$ is an abelian group. Then the universal coefficient theorem relates homology for trivial group actions of $G$ on $\mathbb{Z}$ and cohomology for trivial group actions of $G$ on $A$. 
The (co)homology for trivial group actions is defined as the (co)homology group $H_{n}(BG,A)$ where $BG$ is the classifying space of $G$. A classifying space of a group $G$ is a topological space whose fundamental group is $G$ and whose universal covering space is at least weakly contractible. With this the universal coefficient theorem for group cohomology relating (co)homology groups with trivial group actions is given by 
\begin{widetext}
\begin{equation}
0\rightarrow Ext(H_{p-1}(G,\mathbb{Z}),A)\rightarrow H^{p}(G,A)\rightarrow Hom_{p}(H_{p}(G,\mathbb{Z}),A)\rightarrow 0 
\end{equation}
\end{widetext}
and the sequence splits (albeit not naturally) and we obtain 
\begin{equation}
H^{p}(G,A)\sim Hom(H_{p}(G,\mathbb{Z}),M)\oplus Ext(H_{p-1}(G,\mathbb{Z}),A)
\end{equation}
Mathematically, the procedure is not as well defined, and clearly I left many aspects undefined. For example, yes, there exists group cohomology with coefficients in elliptic curves, but there may not exist a universal coefficient theorem linking such (co)homologies. Point being, mathematics is not yet as developed to make any statements regarding that aspect in general. However, going back to the constructions of physics, more interesting things can be said. 
\par It is interesting how the fact that quantum mechanics had access to global data imposed the mass conservation of the Galilei group [8]. In order to see this, I will closely follow the well known results of [13].  Taking into account the infinitesimal elements of the one-parameter subgroups of the Galilei group considered as a Lie group and using the group law 
\begin{widetext}
\begin{equation}
G'G = (b',a',v',R')(b,a,v,R)=(b'+b,a'+R'a+bv', v'+R'v, R'R)
\end{equation}
\end{widetext}
we can determine their commutators and form the Lie brackets for the Lie algebra of the group. We take the basis elements of the algebra being $\tau$ the time translations, $k_{i}(i=1,2,3)$ the space translations, $u_{i}(i=1,2,3)$ the pure Galilei transformations, and $M_{i}(i=1,2,3)$ the rotations. We get then 
\begin{equation}
[M_{i},M_{j}]=\epsilon_{ijk}M_{k},\;\; [u_{i},u_{j}]=[k_{i},k_{j}]=0
\end{equation}
\begin{equation}
[M_{i},u_{j}]=\epsilon_{ijk}u_{k},\;\; [k_{i},\tau]=0
\end{equation}
\begin{equation}
[M_{i},k_{j}]=\epsilon_{ijk}k_{k},\;\; [u_{i},k_{j}]=0
\end{equation}
\begin{equation}
[M_{i},\tau]=0,\;\; [u_{i},\tau]=k_{i}
\end{equation}
We have the physical representation of the group in the form of 
\begin{widetext}
\begin{equation}
U(b,a,v,R)\psi(p,E,\xi)=exp[-i\frac{1}{2}m a\cdot v+i a\cdot p'-ibE']\sum_{\xi}\psi(p',E',\xi)[D^{s}(R)]_{\xi\zeta}
\end{equation}
\end{widetext}
where $\xi,\; \zeta$ refer to the spin degrees of freedom and 
\begin{equation}
\begin{array}{c}
p'=Rp+mv\\
E'=E+v\cdot Rp +\frac{1}{2}mv^{2}
\end{array}
\end{equation}
where $D^{s}$ characterises the particle behaviour with respect to rotations. We can write with respect to this the infinitesimal transformations as
\begin{equation}
\begin{array}{c}
M=-p\times (\frac{\partial}{\partial p})-iS\\
u=m(\frac{\partial}{\partial p})+p(\frac{\partial}{\partial E})\\
k=p, \;\; \tau=E\\
\end{array}
\end{equation}
This realisation adds a new commutation relation, in the sense that now the translations and pure Galilei transformations no longer commute
\begin{equation}
[u_{i},k_{j}]=m\delta_{ij}
\end{equation}
What we have in fact is a projective representation. We therefore obtained a representation of the Lie algebra of a central extension of the Galilei group and not of the Galilei group itself. The Lie algebra element of the one-parameter subgroup by which the extension was made is $\mu=m$. The extension is central so that $\mu$ commutes with all other Lie algebra elements but it is not trivial so that $\mu$ must appear in the Lie brackets. The enveloping algebra we can form now admits as invariants 
\begin{equation}
\begin{array}{c}
2\mu\tau-k^{2}=2mE-p^{2}=2m\mathcal{V}\\
-(\mu M+k\times u)^{2}=m^{2}S^{2}=m^{2}s(s+1)\\
\mu=m\\
\end{array}
\end{equation}
We conclude that the Galilei group allows us to consider the internal energy of an isolated particle as an arbitrary parameter. In particular all representations $(m,\mathcal{V},s)$ and $(m,\mathcal{V}',s)$ are physically equivalent as shown already in ref [13]. However, we see that $\mathcal{V}$ is an element of the centre of the group algebra. Therefore an equivalence transformation can modify this centre. The observation we need to make is that we work with an extension of the Galilei group and such an extension doesn't have a unique Lie algebra. We have in fact a class of algebras in one-to-one correspondence with the distinct but equivalent systems of factors of the projective representation associated with the extension [13]. When we go from one algebra to another equivalent one, we have to modify the centre element $\mathcal{V}$ only. The fact that within the centre of the enveloping algebra of the group we find a basis element of this algebra (in this case $\mu$) means that we have a superselection rule. If we have a wavefunction formed from the superposition of two states of different mass, again referring to [13],
\begin{equation}
\psi=\psi_{1}+\psi_{2}
\end{equation}
with the two $\psi_{1}$, $\psi_{2}$ belonging to the physical representations $(m_{1}, \mathcal{V},s)$ and 
$(m_{2},\mathcal{V},s)$ of the Galilei group then we can transform this composed state under the following series of transformations: a translation $a$, a pure Galilei transformation $v$, the inverse translation, and the inverse Galilei transformation
\begin{equation}
(0,0,-v,1)(0,-a,0,1)(0,0,v,1)(0,a,0,1)=(0,0,0,1)
\end{equation}
and we get the identical transformation. However, with respect to some physical representation this set of operations is represented by a phase factor at most. 
\begin{widetext}
\begin{equation}
U(0,0,-v,1)U(0,-a,0,1)U(0,0,v,1)U(0,a,0,1)=e^{-ima\cdot v}
\end{equation}
\end{widetext}
and the composite state becomes 
\begin{equation}
\psi=\psi_{1}+\psi_{2}\rightarrow \psi=e^{-im_{1}a\cdot v}\psi_{1}+e^{-im_{2}a\cdot v}\psi_{2}
\end{equation}
If $m_{1}\neq m_{2}$ then an identical transformation would affect the norm of any of the component states. The relative phase of the two states having different masses is arbitrary and therefore quantum mechanics should not allow states with a mass spectrum or unstable particles. I had to explain this in detail to make it clear how quantum mechanics can detect features of the group we use and how superpositions are affected by the type of groups we choose. 
\par Let us see if we can understand the transition between different theories now in terms of quantum entanglement and non-separability. As we have seen above, the change in the cohomology of the underlying symmetry group associated to quantum mechanics eliminates a superselection rule imposed by the group cohomology of the Galilei group. In that sense, we see this change in connectivity and the reduction of the topological obstruction in the symmetry group as a "wormhole" which affects the types of superpositions we can form in quantum mechanics. Indeed, in the realistic situation in which we have local Lorentz symmetry the group has trivial cohomology and we can indeed form states with a spectrum of masses. This allows the decay of particles, as well as the creation of entanglement between the decay products. So, we have seen that the "wormhole is entanglement" idea can have an interpretation in other structures, for example in groups and their topological structure, as opposed to only thinking in terms of boundary and/or bulk spaces. This brings us to the situation of dualities in string theory. 
\par It is interesting to see how in string theory the quantisation is being performed. The main idea of introducing a string is to extend the particle to an object that has an additional dimension. This action cannot exist all by itself, as there are a series of symmetries that need to be obeyed for the theory to properly take into account the principles of physics we already know. Among those, of course we start with local Lorentz invariance and diffeomorphism invariance to which we add the Weyl invariance on the string worldsheet. Quantisation however takes into account the global structure of any underlying manifold, and because of that not all symmetries are compatible with each other. Moreover, it is not always simple to lift the diffeomorphism operation to quantum mechanics. To start from the very beginning let us introduce the Witt algebra as 
\begin{equation}
g=\mathbb{R}[L_{n}:n\in\mathbb{Z}]
\end{equation}
where $L_{n}$ is the operator on the Laurent polynomial ring $\mathbb{R}[z,z^{-1}]$ given by $L_{n} p(z)=z^{1-n}p'(z)$. The Witt algebra then is being generated by such elements $(L_{n})_{n\in\mathbb{Z}}$ that satisfy the commutation relations 
\begin{equation}
[L_{m},L_{n}]=(m-n)L_{m+n}, \;\; m,n\in \mathbb{Z}
\end{equation}
If we replace $\mathbb{R}$ with $\mathbb{C}$ we define our algebra now $g_{\mathbb{C}}$. The group of diffeomorphisms on a smooth manifold $M$, $Diff(M)$ has a Lie group structure and we then have a Lie algebra of this group $Vect(M)$ consisting of the smooth vector fields on $M$. If $M$ is chosen to be the unit circle $\mathbb{S}^{1}$ we obtain a Lie group that has a Lie algebra defined by the Witt algebra. To see this we look at an infinitesimal diffeomorphism on the circle and identify it to a vector field on $\mathbb{S}^{1}$. The vector field is for example a field of tangent vectors. Each tangent vector can be seen as a multiple of $\partial_{\theta}$ so the vector field can be described by $A(\theta)\partial_{\theta}$ where $A(\theta)$ are smooth functions on $\mathbb{S}^{1}$. We can rewrite $A(\theta)$ as a Fourier series and the space of vector fields is generated by $\{e^{i n \theta}|n\in\mathbb{Z}\}$. 
Now let $L_{n}=-ie^{i n \theta}\partial_{\theta}$. These satisfy the generator relations for the Witt algebra for all smooth functions $f$ on $\mathbb{S}^{1}$
\begin{equation}
[L_{n},L_{m}]f=e^{i(n+m)\theta}i(n-m)f=(m-n)L_{n+m}f
\end{equation}
which leads to the well known Witt algebra relations
\begin{equation}
[L_{n},L_{m}]=(m-n)L_{n+m}
\end{equation}
This algebra is describing the operations associated to diffeomorphism invariance for our $S^{1}$ object and is the basis of the generation of the gauge invariance of a (closed bosonic, in this simple example) string theory. In any case, quantisation implies lifting the theory to a recovery of the de-projectivised representation of the structural symmetry group. We can imagine the Witt algebra acting on the Heisenberg algebra by derivations, but in fact at the level of representations, the Witt algebra does not act on the Fock representations directly. The usual solution to this is the construction of a central extension of the Witt algebra by co-cycle. Bringing together the Lorentz/Poincare symmetry, the Diffeomorphism invariance, and the Weyl symmetry in a quantised theory however is not quite trivial. Of course, we have to go to the central extension to make sense of the action of our operators on the Fock space we construct, but even so, we encounter anomalies. Namely, we notice that the central extension of the Witt algebra in which we work, deviates from the algebra of the operators we would like. If we consider the stress energy tensor, under conformal transformations it will gain a term that violates the tensor law, and in that way we obtain a central charge of the transformation. The central charge is precisely the amount by which the stress energy tensor violates the tensor law at the classical level. If we continue with a quantisation prescription, say BRST we obtain an additional degree of freedom that violates Lorentz invariance. This is a result of the fact that in Lorentz symmetry is not quite compatible with the string symmetries under quantisation. However, we can solve this problem by adding ghost fields in the process of the BRST quantisation. Indeed, we have to fix three gauge symmetries in string theory, and we do that by promoting the Jacobian determinants obtained via the gauge fixing conditions under the path integrals to a field dependent expression using the relation between the determinant and the exponential for ghost fields. The Jacobian is simply the best linear approximation of the area resulting from a deformation of the integration surface by a given transformation. The transformation here is given by the gauge fixing condition, which then the field representation of the Jacobian transports consistently across the entire manifold. However, by doing this, aside of fixing the gauge (or making a gauge choice) one also expands the Fock space to a Fock space that contains in essence also ghost mode operators. The ghost modes are extremely important in this context as they implement the gauge choice as a variable transformation that must be consistently transported throughout our space. This ghost system will contribute to the stress energy tensor allowing us to rewrite it in a covariant way but will also deform our algebra. A particular choice of the coefficients appearing in the ghost corrections obtained in our algebra allows us to then recover the Witt expression for the algebra
\begin{equation}
[L_{m}, L_{n}]=(m-n)L_{m+n}
\end{equation}
for our quantised theory, however, on a larger Fock space that includes the ghost states. Physical states are to be associated to BRST cohomology classes. But for that we need the BRST charge operator, which is the equivalent of the exterior derivative in a cohomology theory. As such, this operator is the equivalent of a boundary operator, and hence, must satisfy the relation $Q^{2}=0$ meaning basically that the boundary of a boundary is nill, or otherwise stated, a true boundary configuration must close upon itself. With this requirement we can show that for a bosonic string, the BRST charge does satisfy this property, leading to a vanishing of the anomaly in the case of D=26 dimensions. This is the basic path that allowed us to obtain a quantised theory of a string (in this case bosonic). We have to remember however some choices that we made in this process. 
We made an exact choice for the group designed for our symmetry to be one that results in the Virasoro algebra locally. We expanded around the Virasoro algebra with various group operators and used a series of ghost fields to make it consistent with quantum mechanics, which in a sense was expected given that it was after all the central extension of the Witt algebra, and that indeed the central extension makes the quantum construction compatible with the symmetries of the extended algebra. However, in all honesty, we do not know what the actual group is supposed to be. The anomaly cancellation that occurs in D=26 is only dependent on us making the assumption that the same symmetries that are valid in non-string cases must be extended by the means of a central, co-cycle based extension, which then leads to a not so surprising BRST cohomology as a domain for our physical states. 
If Einstein were to think in a similar way, lacking additional experimental evidence from, say, Maxwell's electrodynamics, we would be working with a central extension of the Galilei group, which amounts to the Bargmann group, but which also includes false, un-natural superselection rules. 
But I am sure it could be argued that there is no factual reason to challenge this approach for string theory, as we know of the actual symmetries, namely local Lorentz invariance, general reparametrisation invariance hence diffeomorphism invariance on the worldsheet, and weyl invariance as a result of a generalisation to a string worldsheet. However, as in the case of the Galilei group, the obstructions appear at the level of quantum mechanics, so, it would make some sense to think of potential modifications originating from our quantum mechanical understanding of Nature. 
\par There was some work done up to now in the field of so called non-critical string theories. There, we assume that we can work with string theories that do not satisfy the cancellation prescriptions for anomalies, and therefore can exist in lower dimensions, say in D=4 dimensions. The problem that appears in such approaches is that the actual Lorentz symmetry or, if not, the various gauge symmetries are lost. Unless we have some strong guiding principle for such a move, there is no justification to make it. 
But is there such a justification?

The main question that must be asked here is one that emerges in quantum information and that is required to unify various aspects of our geometrical understanding of Nature. The question is "how do we know something about somewhere else?" The Lorentz group of transformations, and implicitly special relativity, are based on the observation that light travels at a constant speed in vacuum, constant in the sense that it doesn't depend on choices of reference frames. However, it is a leap of faith to assume that the extraction of information about an event that is separated in space must only be performed by means of light signals. In fact, quantum information is based on the fact that the amount of information that can be transmitted by classical means (that means, by light pulses) can be drastically diminished if there are some types of correlations that involve global, non-separable information shared by the two participants. Our concepts of spacetime intervals are defined by means of classical light rays being the sole source of information retrieval. This leads us to the Lorentz group. However, if entanglement is involved, some conjectures claim that the geometry of spacetime is altered (say ER=EPR). We can understand how that would affect large black holes entangled with each other, having wormhole geometries between them, but how would that affect the local structure of spacetime, and in particular the Lorentz group transformations? Quantum mechanics allows us to probe into the non-perturbative and topological regimes of our underlying manifolds and our group structures. The existence of anomalies in the quantisation of a string when the Lorentz symmetry is put together with the Weyl and reparametrisation symmetries of the worldsheet could mean other things than the fact that we need to expand the number of dimensions of spacetime. One might ask what kind of modifications to the Lorentz group should be made in order to take into account that information retrieval "speed" limitations due to quantum mechanics must be superposed with the light transmission limit of the traditional special relativity? 
It is well known that the Lie algebra is a local approximation of a Lie group, and that in principle several possible Lie groups can be associated to the same Lie algebra. In fact, Lie groups that are isomorphic to each other have isomorphic Lie algebras, but the reverse is not true. Only if we consider simply connected Lie groups do we have the correspondence between Lie algebras and Lie groups one-to-one. We can have isomorphic Lie algebras that correspond to different Lie groups.  Quantum mechanics should in principle be sensitive to the "true" Lie group of Nature, and therefore we should see discrepancies once situations are involved where quantum mechanics may "probe" into this natural group that deviates from our expectations. It is possible that the anomalies to the quantisation of string theory suggest deviations from the general group laws we imagined previously when considering only the local Lie algebraic realisations. It does not seem that those deformations are only visible in the high energy domain of string theory. It seems like the relatively low energy region around a black hole horizon already makes many such problems manifest. 
The transition between symmetry groups and some form of extension occurs in physics in various situations. Wigner's theorem states that a symmetry of a quantum mechanical system determines an (anti-) unitary transformation in the Hilbert space up to a phase factor of the form $e^{i\theta}$. Given therefore a symmetry group $G$ there exists an extension $G'$ of $G$ by $U(1)$ which acts like a group of unitary transformations on the Hilbert space. However, most of the time, such central extensions are not directly used, as larger symmetry groups are usually desirable. For example we do not use the group $SO(3)$ as often, simply because it has the tendency to produce a sign ambiguity, and we can use its universal cover $SU(2)$ where the sign disappears. In the same way, the connected component of the Lorentz group has a similar problem which is avoided by using the universal covering $SL(2, \mathbb{C})$. This however does not work in the case of the translation group $\mathbb{R}^{2n}$ of translations in both positions and momenta, where the phase factors cannot be hidden. One therefore has to work with the central extension, which is the Heisenberg group. As showed previously, the same happens for the Galilei group which is not a symmetry of the Schrodinger equations but its central extension, the Bargmann group is. Another example worth mentioning is of course the algebra of currents, which is related to the Kac-Moody algebras which are themselves the universal central extensions of the loop algebras. It seems suggestive that in both situations in which a central extension was used to introduce a gauge symmetry in quantum mechanics, there was something fundamentally new going on in physics. Galilei had to be replaced with Lorentz, and the anomaly of lifting Galilei to a quantum symmetry was eliminated, and the phase space translation $R^{2n}$ with its non-trivial phase was the indication for quantum mechanics. The combination of Lorentz, Diffeomorphisms, and Weyl and the resulting string anomalies could be the indication of some new physics as well. 
Given our function $c(g,h)$ encoding the non-trivial phase above, $\phi:G\times G\rightarrow U(1)$ such that $U(gh)=U(g)\cdot U(h)\cdot c(g,h)$, we notice that $U(g)$ can be chosen such that $c(g,h)=1$ then $U$ is a unitary representation of $G$ on the Hilbert space. If that is not the case, we have a projective (or ray) representation because the phase factor $c$ keeps appearing. Using this $c$ we can construct a unitary central extension $G'$ of $G$. It has been observed that when $G$ is semi-simple and simply connected as a Lie group, we can choose $U(g)$ such that $c(g,h)=1$ as a constant. Then $G$ can be represented as a symmetry group on the Hilbert space. Looking at the previous examples, the group SO(3) is semi-simple but is not simply connected, therefore we have an ambiguity $c(g,h)=\pm 1$ and hence we can form a universal cover $SU(2)$ which is semi-simple and simply connected and is a symmetry group on the Hilbert space. The restricted Lorentz group is semi-simple but not simply connected, therefore we have a double covering $SL(2,\mathbb{C})$ which is semi-simple and simply connected and a symmetry on the Hilbert space. The group of phase space translations $\mathbb{R}^{2n}$ acting as translations of positions and momenta is indeed simply connected, but it is not semi-simple, therefore we have to go to a central extension which is the Heisenberg group. The same thing happens with the Galilei group which is neither semi-simple nor simply connected and requires a central extension which is the Bragmann group. We see that we indeed finally had to modify the theories we had before in the cases in which the compatibility with quantum mechanics was violated by the group being either simply connected but not semi-simple or neither simply connected nor semi-simple. 
The fact that a group has a non-semi-simple structure means that it cannot be decomposed into irreducible (simple) components, or that the group itself is not such an irreducible component. Basically a non-semi-simple group is one in which there is interference between its parts. 


\section{Localisation and foundations of string theory}
The Lorentz group and the theory of relativity have been introduced via an imaginary experiment in which the causal structure of spacetime was determined by means of light beams and clocks interacting with each other at certain structures called "events" or "spacetime points". The notion of absolute position has of course been abandoned with the observation that general relativity can only have gauge invariant observables that are to some extent at least non-local. Local gauge invariant gravitational observables are not defined. Of course, approximations can be constructed in which one or the other of the properties becomes irrelevant given a certain context, but in absolute terms, the observation above remains correct. In any case, it is often forgotten that Einstein's Gedankenexperiment worked with light beams (or some zero mass gauge fields, to use modern terminology) and their interactions at specific points. The causal structure of spacetime as well as the existence of the notion of "spacetime events" was linked to such an interaction occurring, by Einstein's definition, "locally". To this concept of locality can be given an absolute sense as long as the spacetime curvature doesn't appear to be important and hence the spacetime is to a decent approximation Minkowski. Once these concepts have been accepted as such, the idea of quantum field theory has been developed, and the standard image of a Feynman diagram became the one in which the associated fields propagate for a while, and then interact in a point, producing an "event" that leads to production and/or annihilation of some particle, given the specific properties of the respective process that we wish to simulate. However, we have to think also backwards: it is this interaction exactly, that allows us to define the causal structure, and hence the Lorentz structure on our manifold. The definition of a Minkowski spacetime, and the idea of interaction of "light beams" (or photons) with matter in well identifiable points are interrelated logically and cannot be separated. 
Let us now re-consider this construction in the case of a string. In fact, the "string" is any structure that we can call "extended", as in, fundamentally not occupying only a single point in spacetime. We can consider it as an "extended event" if we wish. This approach makes indeed more sense, as the idea that interaction occurs only in a point-like region is more of a reductionist abstraction than something that could even in principle be measured in reality. We could therefore replace the "light beam" or "photons" of Einstein's Gedankenexperiment with strings or extended objects and see what kind of causal structure that would produce. Constructing a gauge theory based on extended objects is not a new idea. In fact, higher gauge theory and higher gauge fields are well known, the fields appearing in such theory are Kalb-Ramond, Ramond-Ramond, the C-fields, and so on. However, the observation here is that if we take one of those extended events and propagate them over a Minkowski spacetime, we cannot possibly have a Minkowski spacetime anymore, at least not with absolute precision. This is not because such objects would obviously leave a spacetime curvature imprint as they travel, which is of course, true (albeit it may be approximated away as a small effect), but simply because our interaction point that would have determined the precise location of a spacetime event, and which would have allowed us to define the causal structure and the Lorentz transformation acting on a point in such a Minkowski spacetime is not absolutely defined anymore. In fact, if we look at the interaction vertex of a string, we can draw curves of constant time (according to our Minkowski spacetime) and determine that two different observers will not agree on the specific point where the splitting of the string (the interaction) actually occurred. For one observer there will be only one string still, for another, there will already be two split strings. As such, this phenomenon is very well known, and doesn't raise any particular mathematical issues if one accepts its common interpretation, and that interpretation is that because of this "extension" of the interaction point, the string theory will not suffer of the same UV divergences as a classical quantum field theory would. This might be true, but in terms of foundational understanding, this creates a series of problems with the construction of a uniquely defined point in spacetime, with the construction of a causal structure (and its meaning once it is constructed) and with the very idea of what a point is supposed to mean on a string. If a string is a fundamental entity, defining a point on a string is an interesting idea. What kind of information can such a point carry, and to what extent is such information even non-ambiguous? Aside of that, how can we re-construct the Minkowski spacetime following Einstein's reasoning of light beams interacting in "points" that represent spacetime events, when we work with strings? Again, mathematically, there is a well known way of dealing with this, leading to the so called "quantisation of a string" but it is not at all a given fact that this is the only possible approach or the correct one. 
Let us think in terms of algebraic geometry, and envisage the situation in which one curve is tangent to another curve. Let us say that the first curve is a field in quantum field theory, and the second curve is another one. The interaction is supposed to occur in a point-like region. Of course, due to renormalisation we know that this is in fact a bit more complicated. We can perform an operator product expansion to deal with the situation of fields defined in the same point. However, in algebraic geometry the situation is explained in a more interesting and different way. In fact, we can consider scheme theory and interpret the first curve tangent to another curve in a point as the ring of polynomial functions on the intersection. We have a ring of polynomial functions on the first curve, and the ring of polynomial functions on the second curve. The ring of polynomial functions on the intersection is the quotient of the ring by both of the associated polynomials. Such a ring will have a non-trivial nilpotent. This nilpotent has the ability to tell us that a function on the scheme that encodes the intersection records the derivative of one curve with respect to the tangent vectors at the intersection point. The two geometric objects share a tangent space. This will bring into the picture the idea that at that point the intersection is not just a point, but in fact two points that are connected by an infinitesimal object which is represented as the nilpotent. 
The intersection "point" is therefore two points, with quite distinct properties. The information related to the processes going on between these two points at the intersection is represented in a perturbative manner in terms of the loop expansion of our field theory. It is definitely a complicated process which may eventually be mapped onto the topological and geometric properties of our infinitesimal object, its cohomological structure, and so on. In string theory, this property is made more manifest, but it doesn't quite represent the actual situation. In a categorical sense, we could restrict the interaction region to a point even in the case of string theory, but then the causal structure should have to be represented in terms of the maps between the respective points occurring sequentially right before and after the interaction. In a sense, Feynman's path integral quantisation does something very similar although it will still require a set of UV "corrections" (or, renormalisation). String theory does not have this problem, but it replaces it with the idea that one interaction point is not uniquely defined in a way that is independent of the choice of reference frame. We have basically two choices here, both of them somewhat expand the idea of the Lorentz group. Either we will obtain a higher dimensional spacetime in which the propagation of our string will take place, and hence the Lorentz group will have to be generalised accordingly when dealing with strings, or, the second alternative, as I interpret it here, we have to look for a higher categorical alternative to the Lorentz group, which will have to deal with the existence of extended objects and ultimately expand the notion of a causal structure to whatever will be there that will encode the causal structure of a "spacetime" with strings. The first path is well known. It has been the subject of research for at least 70 years now but it has some obvious problems that are not of procedural mathematical nature. The problem is that when producing a string structure in this way, using the same generalisation of symmetry as in standard string theory, we forget much of the categorical information that would be available in a full interpretation of the "symmetry group" involved in extended objects. In a sense, we perform a partial decategorification. What happens then is that when constructing string theory, we claim to perform a full categorification, but when applying Lorentz symmetry, and then gauge symmetry, we partially (and unintentionally) decategorify our construction to some intermediary stage that has a dubious interpretation.
This conceptual difficulty leads on one side to the requirement of renormalisation in quantum field theories, which is in principle a decent mathematical prescription, which, according to my interpretation, continues the decategorification to its ultimate conclusion, resulting in effective theories that completely forget some essential information from the UV cutoff or beyond (see ref. [14]), or, we preserve a partial categorification but continue with a string theory with dubious interpretations and an unclear notion of Minkowski spacetime in which the "spacetime events" are defined as being fundamentally ambiguous. In that sense, the prescription of Einstein's Gedankenexperiment is not valid and it is of no real surprise that we obtain violations of Lorentz symmetry in various effective theories emerging from this worldview. This is not to say that such violations may not have a physical underpinning. I do not know whether they do or not. In the literature, some Lorentz violations have been eliminated by considering to some extent non-perturbative effects, which however can be linked to the inclusion of more categorical maps, in the sense I explained above. It would however be interesting to construct a higher categorical interpretation of the Lorentz group in which extended objects could be dealt with in a consistent way. I assume that in this sense, one of the categorifications of geometry available, known as Synthetic differential geometry could be of use. 
The distinction that I needed to clarify here is that between the interpretation of a Lorentz invariant string and a Lorentz invariant fundamental string. Of course, imagining a string-like object in spacetime in which Lorentz symmetry is respected locally should not be difficult. However, this would not be a fundamental string that defines the types of interactions possible at a fundamental level. A fundamental string should be at the origin of the causal structure, which is constructed by replacing the "beam of light" Gedankenexperiment of Einstein, by a "beam of strings" experiment where the "events" are multi-valued and Lorentz group "ambiguous". It seems like string theory constructed the quantisation of a string by considering the fundamental string as a common string inside spacetime, that would have nothing to do with the construction of the causal structure, and hence of spacetime, which, a truly fundamental string at the origin of all interactions would have had to have. 
From this point of view, string theory as constructed today seems rather naive. It claims to rely on a very fundamental type of extended object but then it constructs a relativistic and even quantum such object which starts from the properties of, literally, a piece of elastic band. However, an elastic band cannot define the structure of spacetime, nor does it lie at the foundations of interactions. The notion string theorists actually wish to talk about is what I call a "fundamental string". However, both the classical relativistic and the quantum theory of strings deals with the string as if it was a simple elastic band that needs to be quantised and made compatible with relativity. 
Because of this, string theory must start from a perturbative construction. It must be perturbative because it starts with an elastic band moving in Minkowski spacetime. However, once we want such an object to play a constructive role in Einstein's Gedankenexperiment that produced the causal structure and the concept of Minkowski spacetime, we have to realise that the process of thinking started somewhat backwards. We talk today about the idea of "spacetime emergence", which is regarded from a point of view that makes the discussion unnecessarily complicated. This is so, because, following Einstein, his causal structure "emerged" from a simple Gedankenexperiment of photons interacting with clocks. If we replace photons with strings, the causal structure won't look the same anymore. If we insist however in using a Minkowski spacetime, as it is done in the first lessons on string theory quantisation, we force our theory to be not only perturbative, but also decategorified. After this task is being done, we are working with a strong approximation of what we actually wanted to construct. This approximation could in principle be generalised in the sense of allowing the spacetime in which the string propagate to have extra dimensions or be curved, but it still doesn't generalise properly to the notion of an extended interaction region, which is why we continue having a perturbative approach and why the "emergence of spacetime" appears as a mystery. 
It is important therefore to remember that Einstein's Gedankenexperiment arrived at the notion of a Lorentz group and of a Minkowski spacetime by the process of considering a very special and reductionist view on what interactions should mean. As a matter of fact, Einstein derived the causal structure which we now use in physics by thinking that events are only determined by interactions between clocks and light beams. Both clocks and light beams are rather rudimentary concepts and they have been replaced by Einstein with points of no dimension. Using this mental idealisation he considered interactions to take place in points and hence events to occur in points. As a logical consequence of this, he derived the causal structure, Minkowski space, general relativity and the Lorentz group. 
String theorists, when developing their theory, constructed a string and introduced the notion of interaction between extended objects. However, they didn't consider that the very idea of interaction plays a fundamental role in the construction of the causal structure. Because of this, they changed the type of interaction into an interaction between extended objects which occurs in an extended domain, but did not consider it to be necessary to update the concepts of causality accordingly, despite the fact that they had as model, Einstein's way of thinking, which, unfortunately they didn't replicate properly. First and foremost, it must be clear that the interaction generated the causal structure, and a point-like interaction forcefully generates a different causal structure than an extended interaction region. In string theory, we force point-wise Lorentz group in a Minkowski spacetime on an extended string. This would work if the string was a usual object which we could assimilate as made up of some inner structure, would contain "points" in it, on which the Lorentz group could act. Such a string however wouldn't qualify as a string if we believe the string theorists' view that such a string should be fundamental, or that it should in principle encode fundamental interactions. A fundamental string has to be incorporated in Einstein's Gedankenexperiment and produce a rule that would be the analogue of the Lorentz group, while surpassing it in various ways. 
Moreover, string theorists construct structure on the string as if it was a mathematical object that would obey standard calculus with the usual definitions of infinitesimals by limits and the ideas of Bolzano-Weierstrass. 
This looks like a strenuous endeavour. In fact, a string should be more accurately defined as an "extended structure", in order to avoid having the impression that one talks about an elastic band made up of smaller elastic structures, "string molecules" so to speak.
The actual reason why strings should be in principle useful is for them to eliminate a part of the unnecessary reductionism in Einstein's approach to the causal structure, namely the assumption that interactions between lights and clocks could actually occur in points. From that point of view, a string should be represented as an extended interaction region, and by this definition, one should try to find how to construct a causal structure where interaction is generalised in that sense. As a note, the classical view of the Lorentz group cannot be valid due to its action on points. A higher categorical expansion is needed.
In a sense, we can say that a non-degenerate metric is defined by Einstein's Gedankenexperiment by means of interaction, but once we consider interaction in all its details, the metric defined by the procedure or Einstein's Gedankenexperiment becomes degenerate. This amounts to an argumentative cycle. The very same thing you use to define your concepts, once you keep going on the argumentation path, it leads you right back to the origin, and alters that origin significantly.  
\section{Categorification of groups}
The approach presented in the previous section has some interesting implications to the idea of categorical renormalisation group presented in [14]. There, the idea of a renormalisation group is extended to a categorical renormalisation group that in principle should be capable of taking into consideration correlations occurring at or beyond the cut-off scale. The categorification of groups opens the possibility of finding not only a relation between different scales via the construction of effective field theories by integrating over degrees of freedom of smaller length scales, but also of finding relations between different such relations and a hierarchisation of such relations. If the maps between different transitions between scales or different methods of producing effective theories are found, and this indeed happens if one uses the categorical renormalisation group, then correlations at or beyond the UV domain have an impact on effective theories that can be constructed at a lower energy. 
However, in the context of the previous section, the problem becomes more interesting. If we think in terms of conformal symmetry of a theory, a renormalisable theory leads to renormalisation group flows of the parameters that contribute with a non-vanishing trace to the energy momentum tensor from the divergence of the scale current 
\begin{equation}
\partial_{\mu}J^{\mu}\sim \beta(g)G_{\alpha\beta}^{a}G^{\alpha\beta a}+...
\end{equation}
However, a categorification of the renormalisation group may be constructed such that not only a hierarchy emerges, but that correlations at the level of the UV cut-off and beyond can be probed and become additional compensatory terms. 
A similar situation occurs in the case of analysing small scale behaviour of interactions, which are constituent concepts in the construction of the Minkowski space. We do not have spacetime points anymore playing the role of events, but instead geometrical regions around classical spacetime "event" points that require a categorical interpretation of the Lorentz group. This implies that the hierarchy problem can be seen as a result of two incorrect approaches: one, in which spacetime points are seen as abstract mathematical points instead of regions on a topos, the second, in which the renormalisation group is seen as a one way transformation between scales instead of a categorical construction that allows a hierarchical structure between the maps that link different scales. 
Special relativity has been built starting from two observations: first, of course, was the invariance of the speed of light at changes of reference frames. Second was the construction of a spacetime in which events were the fundamental points and such events were determined by the interactions, in particular the interactions of light with massive particles. Those interactions were assumed to occur in specific geometric points, objects with zero dimension distributed according to the connectivity of the real numbers. 
This however is an approximation. Consider for example the point described by a real line starting from the centre of a page towards the right, located at 1.53 units of length, say, meters, to the right. Can that indeed be a point in spacetime? In order for this to be possible, one has to have an interaction marking precisely that point as a "spacetime event". To probe such an event one has to send a photon towards that point. However, if the energy of the photon is such that its associated wavelength is smaller or equal to 1.2132 picometers or $1.2132\times 10^{-12}$ meters a pair production will occur, and, if an electron was indeed at 1.53 meters to the right at the given location, another pair of electrons and positrons will be produced way before one reaches $1.53\pm 1.2132\times 10^{-12}$ meters. The event at 1.53 meters cannot be regarded as a single event, or at least one cannot precisely localise the event at that point on the real line, because in the process one generates a series of other events around it. If one really wishes to localise an event in order to determine "points" in spacetime, one has to remember that such points are not geometric points in some real-values manifold, but in fact equivalence classes in which different events can be regarded as distinct while not truly distinguishable from the actual event one has to consider. The logic of such a geometry is definitely not classical, as having a statement saying that the event "is" indeed ocurring at 1.53 meters on the real line does not deny the possibility that the event is occurring at 1.2132 picometers away. Therefore, denying that the event is occurring at 1.53 meters on the real line implies a series of events occurring in a sphere of radius 1.2132 picometers around the point at 1.53 meters, that are clearly distinct from the perspective of the real line, hence are not the event at 1.53 meters, but are impossible to be discerned from it if one wishes to describe the event as an "interaction", therefore they are not the event, but not significantly distinct from the event either. This structure reminds us of the intuitionistic logic in Topos theory which I will discuss in the next chapter. 
Such constructions have been used to argue for a series of ideas about spacetime, for example it has been proposed that spacetime is "fluctuating" or that it is even "discrete". I do not believe such classifications are in fact meaningful. In both cases we assume that spacetime is already there, it has a certain abstract realisation, and it is something beyond what could in principle be measured. However, I believe Einstein's view was the most insightful regarding the nature of spacetime. He saw it as a collection of events, as only the events are in fact measurable. Such events, appearing as interactions that can determine that something is going on somewhere, are the foundations of the construction of the spacetime manifolds. The only departure from Einstein is that the events themselves cannot be regarded as geometric points or numbers in the real field. If one attempts to isolate one number on the real field with an event, one will immediately notice that a series of other numbers will unavoidably be associated to it, the event being never truly localised. We noticed that this happens in the case of string interactions in the previous chapter. Another extreme limit is the case in which enough quantum field theoretical modes are excited in a region for a black hole and a horizon to be formed. But such are extreme situations of a phenomenon that actually occurs in any situation in which interaction is present. A consequence of this is the requirement of renormalisation and of the construction of a renormalisation group. However, in all those constructions, the nature of the underlying Minkowski space was not noticed to be affected by the new understanding of the idea of interaction. In fact, the requirement of renormalisation of interactions plays a fundamental role in the definition of the various invariants that can be introduced and therefore alter the concepts of events and of spacetime intervals, as well as that of Lorentz transformations. If the connections via light were realised between abstract points on an abstract spacetime, then the relations would be the ones we know. However, the connection via light is realised between spacetime events, which are regions of interaction, and those regions must rely on an intuitionistic geometry encoded in a Topos. 
The advantage of a Topos is manifold. In a Topos construction, the logic associated with the underlying geometry is not classical but usually intuitionistic. This amounts to the idea that the negation of a negation is not necessarily an affirmation, but that it stays in a state of truth that is somewhat different from both the affirmation and of the negation. In this context, the notion of an inverse is weakened as a mathematical concept, and hence a group structure can exist in which the inverse is not the usual inverse, but a weakened form of inverse associated to the possible existence of this third state of truth. 
The basic idea is that the notion of an interaction is what defines the causal structure due to the interaction being at the foundation of the definition of a spacetime event. In Einstein's view, the causal structure and the events in spacetime were literally defined as geometric points where point-like particles interact in the sense of collisions. Einstein defined his events as those spacetime points where a light beam collides with a massive clock. After a reductionist process, those objects became a massless gauge field and a massive point like particle. Needless to say that this represents a reduction, but if we want to know precisely what such an interaction region is, we will notice we don't quite know. In a perturbative sense we can perform some form of calculations by constructing a perturbative series of Feynman diagrams in which the inner "propagators" are replaced with inner loops which then have to be renormalised in a standard fashion. While the inner loops of a perturbative expansion of Feynman diagrams are indeed quantum in nature, the virtual loops of particles being clearly quantum structures, the same concept cannot be extended to spacetime as is the case with what we call now "spacetime fluctuations". The concept of fluctuations appears due to an operator associated with an observable, that has a matrix representation, which acts on a wavefunction, producing each time a different eigenvalue. Statistically, after repeating the measurements several times, we will notice a fluctuation simply due to the fact that each eigenvalue occurs with a certain probability and that will be visible in our distribution. However, the phenomenon occurring with spacetime is very different. When dealing with operator product expansions, we assume our distributions are not directly well defined when the limits $t_{1}\rightarrow t_{2}$ and $x_{1}\rightarrow x_{2}$ are taken, but we assume that the limits can be taken due to the various topological properties of our Minkowski space which we associate to some form of expansion of the real line $\mathbb{R}^{(1,2)}$. The limit of the spacetime points can be taken, while the distributions on spacetime may be ill defined. 
In the case I discuss here, there is no unique point where $t_{1}\rightarrow t_{2}$ or by that matter $x_{1}\rightarrow x_{2}$. The reason for that is the fact that our causal structure and our spacetime is defined as a set of points which represent the region of interaction itself. The spacetime events are not associated to points, in a physical sense, but to regions of interaction. If we can approximate those regions with points in which particles bounce off each other in a cvasi-Newtonian sense, this picture makes sense. However, this is not at all what happens. When we construct quantum field theory we still construct it on Minkowski spacetime, but the resulting interactions between fields or particles defy the exact definition of such a Minkowski spacetime. 
One may think that the "regions" of interactions are regions of some spacetime which we can in principle construct and for which we may define a metric. Mathematically that is possible only if we assume the existence of such a spacetime in an axiomatic sense. If we abandon such axioms or postulates the "region" is not, in fact, a region of spacetime, but in fact, only the definition of an event. Therefore, it would not be sufficient to think in terms of a discretisation of spacetime or in terms of spacetime as being covered by elementary "tiles". In both cases, we would assume some geometric structure before we actually define, from a physical standpoint, what such a geometric structure would be. 
The solution to this dilemma that I propose is to adopt a categorical viewpoint and to introduce the notion of a Topos, while denying the law of excluded middle in the formulation of our geometry. 
In fact, in Topos theory and further in synthetic geometry the inverse operation has a spectrum of strengths supported by theorems introduced by [17] and then [18]-[20]. 
But inversion plays an interesting role in the very first attempts to construct a quantised gauge theory. We can have a brief look at the construction of simple QED 
\begin{equation}
\begin{array}{c}
Z[J]=\int \mathcal{D}A exp\{i S[A]+i\int J^{\mu}A_{\mu}\}=\\
\\
=\int \mathcal{D}A exp\{i \int d^{4}x \{\frac{1}{4}F^{\mu\nu}F_{\mu\nu}+J^{\mu}A_{\mu}\}\}
\end{array}
\end{equation}
If we go to momentum space 
\begin{equation}
A_{\mu}(x)\int\frac{d^{4}p}{(2\pi)^{4}}e^{ip\cdot x}A_{\mu}(p)
\end{equation}
one can re-write the term in the exponential as 
\begin{widetext}
\begin{equation}
\begin{array}{c}
\int_{x}\{-\frac{1}{4}F^{\mu\nu}F_{\mu\nu}+J^{\mu}A_{\mu}\}=\\
\\
=\frac{1}{2}\int \frac{d^{4}p}{(2\pi)^{4}}\{-A_{\mu}(-p)(p^{2}\eta^{\mu\nu}-p^{\mu}p^{\nu})A_{\nu}(p)+J^{\mu}(-p)A_{\mu}(p)+A_{\mu}(-p)J^{\mu}(p)\}
\end{array}
\end{equation}
\end{widetext}
We need to perform a gaussian integral over the gauge field $A_{\mu}$ to perform quantisation but the propagator for the gauge field is not invertible.
in fact we can write it as 
\begin{equation}
\mathcal{P}_{\mu}^{\;\;\nu}(p)=\delta_{\mu}^{\;\;\nu}-\frac{p_{\mu}p^{\nu}}{p^{2}}
\end{equation}
which is a projector on the space orthogonal to $p_{\nu}$. This projector matrix should have eigenvalues 0 or 1, but the projector acting on the momentum is 
\begin{equation}
\mathcal{P}_{\mu}^{\;\;\nu}(p)p_{\nu}=0
\end{equation}
We can therefore decompose the fields as 
\begin{equation}
A_{\nu}(p)=\frac{i}{e}p_{\nu}\beta(p)+\mathcal{P}_{\nu}^{\;\;\rho}(p)A_{\rho}(p)
\end{equation}
such that $p^{\nu}\mathcal{P}_{\nu}^{\;\;\rho}(p)A_{\rho}(p)=0$. 
We see that 
\begin{equation}
\beta(p)=\frac{e}{ip^{2}}p^{\nu}A_{\nu}(p)
\end{equation}
The gauge field however has a gauge symmetry and it can be changed by means of a position or momentum gauge shift. In momentum space we have
\begin{equation}
A_{\mu}(p)\rightarrow \frac{i}{e}p_{\mu}\alpha(p)+A_{\mu}(p)
\end{equation}
Such a gauge transformation can be used to get to $\beta(p)=0$ which leads to 
\begin{equation}
\partial^{\mu}A_{\mu}(x)=0
\end{equation}
With this gauge fixing one can then obtain the inverse propagator and produce the functional Gaussian integral (completing the squares, etc.) 
\begin{widetext}
\begin{equation}
\begin{array}{c}
Z[J]=\int \mathcal{D}A exp\{\frac{i}{2}\int_{p}[-(A_{\mu}(-p)-J_{\rho}(-p)\frac{\mathcal{P}^{\rho}_{\;\;\mu}}{p^{2}})p^{2}\mathcal{P}^{\mu\nu}(A_{\nu}(p)-\frac{\mathcal{P}_{\nu}^{\;\;\sigma}}{p^{2}}J_{\sigma}(p))]\}\times \\
\\
\times exp[\frac{i}{2}\int_{p}J^{\mu}(-p)\frac{\mathcal{P}_{\mu\nu}(p)}{p^{2}}J^{\nu}(p)]=\\
\\
=exp[\frac{i}{2}\int_{x,y}J^{\mu}(x)\Delta_{\mu\nu}(x-y)J^{\nu}(y)]
\end{array}
\end{equation}
\end{widetext}
where the propagator for the photon, after choosing the above gauge is 
\begin{equation}
\Delta_{\mu\nu}(x-y)=\int\frac{d^{4}p}{(2\pi)^{4}}e^{ip(x-y)}\frac{\mathcal{P}_{\mu\nu}(p)}{p^{2}-i\epsilon}
\end{equation}
This is a standard textbook derivation of the propagator in quantum electrodynamics. While everything should be clear in this calculation, the existence of gauge degrees of freedom makes the inversion procedure hard, and in fact, the local gauge structure makes it ill defined. We integrate in the final step only over the configurations that satisfy the gauge condition. In the context of intuitionistic logic, the inverse can have various degrees of precision, as a one-to-one relation between the various points around our reference point that have an "intermediary" truth value, and whatever point outside is not fully defined. Therefore, our definition of the underlying geometry must take this into account. 
The non-abelian case is a generalisation of the above case for the situation in which the gauge generators do not commute. In this case the determinant of the derivative of the Gauge fixing function with respect to the gauge parameters does not depend on the gauge parameters but it does, as opposed to the Abelian case, depend on the gauge fields. Therefore the gauge fixing procedure is affected by the specific region of gauge field integration. In this case, the determinant is associated to the integration of fermionic degrees of freedom which transform like scalars (hence like bosons) under Lorentz transformation. Such fields are the Fadeev-Popov ghosts, and their role is to make consistent the Gauge choice across the field space. This works to some extent, of course. The Gribov ambiguity and other problems emerging in non-abelian gauge theory however are beyond the scope of this article. Needless to say, ghosts, anti-ghosts, fields and anti-fields can be interpreted in more complicated situations and appear as requirements for the closure of the Lie algebra associated to supergravity theories, etc. I see no reason to go into this in detail in this article and therefore I will direct the reader towards [21].
More interestingly, the observation I make in this article has an important impact on the definition of entanglement entropy in the context of string theory. 
It is well known that in quantum field theories, the construction of an entanglement entropy is difficult mainly due to the existence of divergences coming from the fact that two regions of a certain spacetime were artificially and drastically separated by some hyperplane while correlations between points infinitesimally close to each other remained strong. While some form of renormalisation is possible, the problem persists in string theory precisely due to the fact that strings do not have a clear point-like structure associated to them. As stated previously the UV-softness of strings and in particular the impossibility to construct a one-to-one function between a point and a region of string interaction results in clear difficulties in defining string entanglement entropy.

The same type of inversion problem that appears in the basic quantisation of gauge theories occurs in a more fundamental way in string theory when trying to construct a meaningful entanglement entropy. The cause of course appears to be somewhat different, namely the difficulty of creating a clear separation of regions in string theory and in general in theories of quantum gravity. This is due to the fact that observables in gravity (and in quantum gravity) are not strictly speaking local, and this non-locality leads to difficulties in creating a strict separation of the Hilbert spaces associated to the two regions. The same problem persists at the level of local operator algebras in string theory, which, by their very name, appear to be unsuited for a problem of quantum gravity. This can be traced back to the problem of non-invertibility as a clear separation would be possible for a clearly invertible function over a manifold. The non-invertibility or the need for a weaker form of invertibility appears precisely because the region around any separating hyperplane contains sufficient correlations to make such an inverse map ill defined, particularly because we enter a region where the Topos structure becomes important and the negation of a negation (seen as a map) is geometrically not bringing us necessarily to the original point. 
A very interesting discussion about the notion of an inverse in Toposes and the various weaker or stronger theorems defining such inverses has been presented in a somewhat overlooked PhD thesis [17].

\section{Synthetic strings}
The problem with string theory and its dualities, in particular the string duality relating small and large objects, is linked with the ultimate definition of "infinitesimal". Physicists should probably know that point-like and infinitesimal do not mean the same thing, and one cannot use the two concepts interchangeably. However, the basis of calculus is such that those two concepts are indeed interchangeable, and in physics we use the same calculus when dealing with Galilei groups, Lorentz groups, or Virasoro groups.

We should however remember (or state) that the Lorentz group transformations appear as a multiplicative formal group law, and that a formal group law is usually employed when an intermediate context must be described, namely in between Lie groups and Lie algebras. 
In general, a Lie group has some form of global structure and is described, albeit not easily, by means of a topological space or a manifold. Lie algebras are a method of linearising Lie groups. They behave like the tangent space to the identity of the group. The Lie algebra is usually enough to determine the connected component of the Lie group. However, the Lie algebra is limited by the fact that it is defined around the identity. In characteristic zero of an underlying ring (meaning zero number of multiplicative identities added together to sum into the additive identity) the Lie algebra is sufficient. However, if the ring is more general, using simply the Lie algebra will not provide us information about all connected components. The Lie algebras simply are not sufficient to probe the entirety of the group. What is required is a formal group, and the associated formal group law, that can probe regions inaccessible to the Lie algebra itself. 
The characteristic of formal group laws is that they can be used in situations in which the objects they act upon are generally higher order infinitesimal. 
As a matter of definition, if we have a commutative ring with identity R, we call a one-parameter formal group $\mathcal{F}$ over $R$ the power series $F(X,Y)\in R[[X,Y]]$ with the properties that $F(X,Y)=X+Y+d_{2}(X,Y)$, where $d_{2}(X,Y)$ are terms of degree larger or equal to $2$, and that $F(X,F(Y,Z))=F(F(X,Y),Z)$, namely the associativity. 
That means there is much more genius in the Lorentz transformations than first assumed. Not only are they adapted to a finite speed of light, which is why they originally have been introduced, but they are also adapted to variations of the infinitesimal objects on which they are applied. The Lorentz group has been in fact constructed in the form of a formal law due to the way it has been expanded from the Galilei group.  However, to make them compatible with quantum mechanics when various other (gauge) symmetries are introduced, one may wish to consider generalisations. As mentioned before, the standard approach is to consider extension, but as I showed before, Nature didn't choose the Bargmann group (the central extension of the Galilei group) but instead the Lorentz group, which is topologically different from the Galilei group and its central extension. However, Nature seems even more subtle than that. As the Lorentz law appears in the form of a formal group law, and is in essence forming a formal multiplicative group by itself, it may be useful to consider it in the context of synthetic differential geometry [9]. Basically, by synthetic geometry we refer to a type of geometry focused on the categorical structure of the spaces involved, and not on the differential description usually employed. 
Using this geometry, it is possible to define a category of partial differential equations, as introduced by A. M. Vinogradov [11], [12].

Because of this, its axiomatic structure is broad enough to encompass generalisations that are not considered in the usual interpretations of geometry, having as ultimate goal the replacement of most if not all differential geometric prescriptions with algebraic ones. While the Lorentz group law is a formal group law, the rest of the gauge symmetries that we require on the string worldsheet are not, in particular, the worldsheet diffeomorphisms and the Weyl transformations. 

\par If we are to follow the standard approach of string theory quantisation, what we do is to implement the string worldsheet gauge symmetries as constraints, and to implement a gauge choice by means of sets of ghosts and anti-ghosts a la BRST cohomology. In a sense the BRST cohomology quantisation prescription uses ghosts as a form of "regularisation" for the potential incompatibility between the Lorentz symmetry and the gauge invariance on the string worldsheet when quantum mechanics is "switched on". We do in fact obtain ghost contributions for example to the stress energy tensor of the string, and if we separate that term out, we notice that together with the matter stress energy tensor we can in fact cancel out the resulting quantum anomalies due to the various gauge symmetries introduced, with the relatively funny requirement that the target or background spacetime is 26-dimensional in the bosonic string case and 10-dimensional in the superstring case. The origin of this high-dimension requirement is therefore to be traced to the way we considered the rest of the gauge symmetries on the worldsheet. One may ask whether something is wrong with them, before one just decides to work in higher dimensional spaces, but this was not, historically, what people did. The fact that the quantum anomalies could be removed only in higher dimensions however led to some ideas of working with so called non-critical string theories, where the conformal symmetry on the worldsheet was simply abandoned. Probably that is not the right way to go. However, there could be other ways to deal with this, aside of the generally accepted compactification of the extra-dimensions. In fact, it could be possible to understand how the quantum anomalies could be eliminated in lower (maybe even (3+1)) dimensions if one tried to express all gauge symmetries in a form more suitable to the extension to extended objects, or in general to objects that have a non-trivial structure. The transition from a point to an extended object is not at all trivial, and in fact it is the subject of synthetic geometry which tries going through a series of intermediate steps (namely the so called infinitesmial spaces of various orders). In fact, we should remember that the Lorentz law is defined locally and acts point-wise. How could it be generalised to see how it would act when the point-object on which it acts is replaced by a string? 
We should not consider the string as "made up" of smaller components. Physicists keep reminding us that we need to see the string as "fundamental", and therefore, although, luckily the Lorentz group appears as a formal group, it would be interesting to see explicitly how it acts when it is not supposed to act on a point-like object but on an extended object? This would amount to a categorification of Lorentz symmetry. But apparently we have it rather easy with the Lorentz symmetry, and not so much with the other gauge symmetries of the worldsheet. Indeed, both diffeo- and Weyl are much more distant from synthetic geometry than Lorentz. Therefore it is a pertinent question to ask whether string theory could be made consistent in lower dimensions provided we gave it proper modifications of the group laws as pertinent to extended objects, and then go to the low energy limit by transitioning to infinitesimal spaces instead of set theoretical point-based spaces. 
In fact, whenever we work with standard calculus, we assume we can take certain limits in which various properties of infinitesimals can be ignored. We can look for example at the notion of a standard derivative. We can assume that, when we calculate a derivative in the limit in which the distance between two positions becomes very small, we can neglect all properties that are of higher order. 
In synthetic differential geometry the properties related to such differentials, in particular the ability to ignore higher powers of infinitesimals, is introduced as an axiom. However, one may ask, if one accepts that some infinitesimal properties can be ignored, what infinitesimal properties cannot be ignored? In fact, it is possible to formulate consistent axiomatic systems in which some special infinitesimal structures cannot be ignored. 
If we assume that our fundamental object is a string (and not a straight line as in synthetic differential geometry) we could ask what kind of fundamental infinitesimal property should we not ignore for string theory to accommodate the usual symmetries on the worldsheet in the usual (3+1) dimensions. This is an alternative way of thinking to the usual compactification approach in string theory. While string theory generally claims to have an extended fundamental object, it doesn't extend the consequences of such an assumption to the differential geometry used to work with string theory or with the symmetry groups that rely on a point-like mathematics. The result is that what we ask in reality to string theory is : if we introduce a tiny vibrating string which we claim to be fundamental, but we continue believing that it is made out of smaller entities (points), which we then use to define various constraints due to the symmetry laws we know from our large scale approach to physics, what do we have to change in the parameters we work with in our point-based physics to accommodate this construction? The answer is in general that we have to extend the dimensions of our space. 
However, the question we should ask is : if we introduce an extended object as a fundamental object, how do we have to change the differential geometry we are customary using, as well as the group laws of the symmetry groups we constructed in our point-based world in order to accommodate such an extended object? This radically changes our perspective on string theory and its meaningful definition in various limits. 

Synthetic geometry is a categorical approach to geometry in which the underlying structure is that of a topos. As such, the logic involved in synthetic geometry is not the classical logic anymore, but instead the so called intuitionistic logic. The main difference between classical logic and intuitionistic logic is the fact that we can in the latter case abandon the rule of excluded middle
\begin{equation}
\phi\lor\neg\phi
\end{equation}
As a result of this convention, the double negation principle is not allowed, and therefore 
\begin{equation}
\neg(\neg \phi)\implies \phi
\end{equation}
is not a valid implication in intuitionistic logic. 
This rule is particularly important in the description of infinitesimals. In general, we say a number is invertible if it is nonzero. However, while in classical logic we can conclude that if a number is non-invertible it must be zero, in intuitionistic logic, this must not be necessarily true. A non-invertible number may be zero, but it may be any other number that may be, for example, in the same equivalence class with zero, given certain properties of that class. 
Another aspect is the notion of linear dependence and independence. In intuitive logic and topos theory those two concepts cannot be directly related to each other in the same way classical logic allows. For example let $M$ be vector space valued in a commutative ring $R$, and let $M$ be in a topos $\mathcal{E}$. We say that an $n$-tuple $v_{1},...,v_{n}\in M$ is linearly free if 
\begin{equation}
\forall \lambda_{1},...,\lambda_{n}\in R, [\bigvee_{i=1}^{n}(\lambda_{i}\#0)\implies \neg(\sum_{i=1}^{n}\lambda_{i}v_{i}=0)]
\end{equation}
is valid in $\mathcal{E}$, where the symbol \# means that the element on the left side is invertible, i.e. it is separated (apart) from zero, or $(\neg(x=0))$. 
We say that the same $n$-tuple is linearly dependent if 
\begin{equation}
\exists \lambda_{1},...,\lambda_{n}\in R, \neg[\bigvee_{i=1}^{n}(\lambda_{i}\#0)\implies \neg(\sum_{i=1}^{n}\lambda_{i}v_{i}=0)]
\end{equation}
is valid in $\mathcal{E}$. 
In intuitive logic the notion of linearly dependent and that of not linearly free are not equivalent. The linearly free vectors $v$ of $M$ are those that can be defined to belong to $\neg\{0\}$ while the non-linearly free vectors belong to $\neg\neg\{0\}$ which however in general is not $\{0\}$. 
Linear independence however, defined in the classical logic sense implies
\begin{equation}
\forall \lambda_{1},...,\lambda_{n}\in R,[\sum_{i=1}^{n}\lambda_{i}v_{i}=0\implies \bigwedge_{i=1}^{n}(\lambda_{i}=0)]
\end{equation}
It is important therefore to notice that linearly free is not equivalent with linearly independent. 

One aspect that looked interesting in an introduction to synthetic differential geometry was the relatively simple observation that when we defined a derivative in calculus, we calculated the first derivative by ignoring the second or higher order variations so, if we had an operator $h$ that generated a variation, we ignored any higher power in it $h^{n}$, $n\leq 2$. When performing compactification we construct a boundary operator, by implementing the requirement that $\delta^{2}=0$, which, in translation from the language of topology means that the boundary of a boundary is zero, or, any boundary must be closed. Of course, not every closed curve is a boundary, but every curve that is a boundary must be closed. These are two instances in which the quantity we ignore, defined as an infinitesimal distance between points in the first case, and as a closing condition upon itself in the second case amount to two types of large scale "accountability" of the null potency of certain operators. In the first case, the fact that our distance is infinitesimal allowed us to ignore pretty much everything about it and only characterise the function at a given point, and in the second case we couldn't ignore quite everything about the object, but we could still take the limit and consider another form of null potency, namely of the boundary operator. 
Synthetic geometry is indeed a very attractive approach to take, but unfortunately it is not fully general. It does introduce an "infinitesimal", the interval $\Delta$, that can be considered a "straight line" and replaces the point, while being an object in the tangent space of our previous manifold, and inside our manifold at the same time. While it is very desirable to have such an extended tangent "point", what we need is something with more structure than just being the extension of a point to a segment. Our string has modes of excitation, we have creation and annihilation operators acting on this fundamental "segment" and therefore this approach is too restrictive. However, some properties that synthetic geometry discovered remain valid. For one, we may find it necessary to think of all the structures that we wish to preserve at the lowest scale and define those synthetic differentials (of order one) in them. The fact that the Lorentz group is a formal group insures that the Lorentz transformations will be preserved when going to a synthetic counterpart (or at least that they can be preserved in first order, would be interesting to see what corrections one could expect to the Lorentz laws if one goes to higher order fundamental differentials). The null potency of the infinitesimal can be found in another part of the quantisation prescription, namely in the BRST quantisation where the BRST charge $Q$ (which is in fact a boundary operator if we think cohomologically) is null potent $(Q^{2}=0)$ and could be seen as part of the infinitesmial topos. The central charge governs the form of the stress energy tensor in the BRST quantisation of the bosonic string and in fact, if we denote a ghost field by $b$ we can see that 
\begin{equation}
\{Q,b\}=T=T^{X}+T^{ghost}
\end{equation}
which means that the anti-commutator of the BRST null-potent charge with a ghost field results in the stress energy tensor and the null-potency of the BRST charge demands that the central charge measuring the conformal anomaly can be "regularised" (that means, properly identified) by the ghosts, and then cancelled by shifting the dimension of the background space to (in the bosonic case) $c=26$. 
This is common knowledge. What is not common knowledge is that while the direct cause for this is the incompatibility of the conformal symmetry with the Lorentz symmetry in a quantum context, the less direct cause of this is the existence of a null-potent differential as interpreted in synthetic geometry. Yes, it is true that the BRST charge is a boundary operator in a cohomology theory, but we can as well see it as a differential null-potent object from a synthetic geometric point of view. Therefore, the deeper cause for the conformal anomaly is the incompatibility of the expected conformal symmetry with the extended structure of the field. This can be compensated in higher dimensions, but it can also be compensated by introducing additional structure to our infinitesimal object, or adding more infinitesimal objects that could solve the problem, while keeping the dimension low. In a sense compactification does that, but not in a very controlled way. The general description is that one compactifies the extra dimensions, therefore introducing a set of closed boundaries for each compactified dimension, say, $\delta_{i}^{2}=0$ for all compactified dimensions $i=1,...,k$ and then one takes the limit in which the radius of compactification is small. Taking the limit is not a trivial operation, and the reason we obtain no consistent low energy descriptions of strings is basically because we take the limit wrong. But that is a discussion for another article. Here I want to focus on alternatives to this operation. Let us see what kind of structure do we need to add in the form of synthetic infinitesimals in order to make the theory compatible to the desired symmetries in lower dimensions. Also, one should be careful to consider that it is not a given that the conformal symmetry needs to be maintained in the same form once extended objects are introduced. After all, it is also interesting to explore what the conformal symmetry would amount to if we were to re-write it in a manifestly formal group sense. 
In any case, let us repeat the BRST quantisation of the bosonic string and see where we have to challenge conventional wisdom. As long as our gauge algebra closes decently fine, and therefore is not directly field dependent we can write our action $S[\phi]$ as a theory that respects the symmetries
\begin{equation}
\delta \phi = \int d\mu_{\alpha}\epsilon^{\alpha}\delta_{\alpha}\phi = \epsilon \delta_{\alpha}\phi
\end{equation}
$\alpha$ simply denotes an index for spacetime coordinates and $d\mu$ is our measure on the $\alpha$-space. The small parameters $\epsilon^{\alpha}$ describe the symmetry. The gauge variation is
\begin{equation}
[\delta_{\alpha},\delta_{\beta}]=f_{\alpha\beta}^{\gamma}\delta_{\gamma}
\end{equation}
and of course, $f_{\alpha\beta}^{\gamma}$ are considered field independent. These structure constants describe the symmetry. For diffeomorphisms we take $\alpha=y$, and a vector $\rho$, so that $\epsilon^{\alpha}\rightarrow \epsilon^{\rho}(y)$ with $d\mu_{\alpha}=d^{d}y$. Given a scalar field, the transformation will act like
\begin{equation}
\delta_{y,\rho}\phi(x)=-\delta^{d}(x-y)\partial_{\rho}\phi(x)
\end{equation}
and we obtain the diffeomorphism $\delta\phi(x)=-\epsilon^{\rho}(x)\partial_{\rho}\phi(x)$. The commutator of two diffeomorphisms on $\phi$ is
\begin{widetext}
\begin{equation}
\begin{array}{c}
[\delta_{y,\rho}, \delta_{z,\sigma}]\phi(x)=\delta^{d}(x-z)\partial_{x^{\sigma}}[\delta^{d}(x-y)\partial_{\rho}\phi(x)]-(y,\rho\leftrightarrow z,\sigma)=\\
\\
=\int d^{d}w[-\delta_{\rho}^{\nu}\delta^{d}(w-z)\partial_{\sigma}\delta^{d}(x-y)+\delta_{\sigma}^{\nu}\delta^{d}(w-y)\partial_{\rho}\delta^{d}(x-z)]\delta_{w,\nu}\phi(x)\\
\end{array}
\end{equation}
\end{widetext}
which gives us the structure constants for diffeomorphisms of the form 
\begin{equation}
f_{(y,\rho)(z,\sigma)}^{(w,\nu)}=-\delta_{\rho}^{\nu}\delta^{d}(w-z)\partial_{\sigma}\delta^{d}(x-y)+\delta_{\sigma}^{\nu}\delta^{d}(w-y)\partial_{\rho}\delta^{d}(x-z)
\end{equation}
The partition function of our theory, that is, its quantum form, will be
\begin{equation}
Z=\int [\mathcal{D}\phi]e^{-S[\phi]}
\end{equation}
The gauge symmetry can be introduced as a constraint $F^{A}[\phi]=0$ where $A$ labels the components of the gauge symmetry. A consistent gauge transformation must also have a non-degenerate Jacobian, which implies that form-wise the transition from one gauge configuration to another linked by a gauge symmetry is well defined. Let us denote the Jacobian of the transformation $\delta_{\alpha}F^{A}[\phi]$ then
\begin{equation}
Z=\int [\mathcal{D}\phi]\delta(F^{A}[\phi])det(\delta_{\alpha}F^{A}[\phi])e^{-S[\phi]}
\end{equation}
This imposes the gauge condition and makes sure that the propagation of the gauge transformation on the coordinates is done consistently, the determinant measuring the linear first order deviation induced by the gauge transformation on the fields. This is an overlooked important point though. 
The Jacobian behaves somewhat like the tangent in simple one dimensional calculus, it is an element of the derivative which we define, as we know since forever, as 
\begin{equation}
f'(x)=\lim_{h\rightarrow 0}\frac{f(x+h)-f(x)}{h}
\end{equation}
In synthetic geometry, the limit when $h\rightarrow 0$ is somewhat different. The first derivative, as expressed in the above simple formula makes the assumption of linearity, namely that we can ignore higher powers of $h$, say $h^{2}$, which means basically that our derivative gives us the best linear approximation of our function in a given point. This is well known, it is the definition of what it means to be a derivative in standard calculus. The Jacobian doesn't do much else, except that it does it for a multivariate context. In fact, the Jacobian for a two form gives us just the best linear two-form approximation of the evaluation of our two form defined manifold in a certain parallelogram region. In gauge space, basically our Jacobian gives us the best local linear approximation of the gauge trajectory, and implements the required correction so that our path integral stays "on path" given the gauge transformation. However, this only works that way if our fundamental object is a point, and hence the limit $h\rightarrow 0$ can be taken in a trivial way. This does not have to be so, and in fact in string theory it is not. So, we had an object that we used in our limit, for which we approximated that $h^{2}=0$. The obvious choice is that of $h\rightarrow 0$, but synthetic geometry teaches us that this is not the only choice. In fact, in synthetic geometry we define our infinitesimal as
\begin{equation}
\Delta=\{x\cdot R|x^{2}=0\}
\end{equation}
as a synthetic differential, which has the property that any object in it, squared, produces zero. $R$ here is not the real line, but the synthetic smooth line. Obviously, we see that here, the limit $h\rightarrow 0$ is not that trivial, and that in fact there are several ways in which we can reach to something that has the same property of being zero when squared. The real reason for that limit was that $0$ was the only element that could have been ignored when taking the square. Synthetic geometry takes away this property. But in fact it is not only synthetic geometry that takes it away. In cohomology, the closed boundary operator has the same property $\delta^{2}=0$.
So, going back to our Jacobian, which is basically a matrix of first order derivatives, we have to remember that the reason why we have chosen it that way was because each of its derivatives gave us the best possible linear approximation in a given point, on a given direction. We also ignored all possible higher order differentials in any of those directions by assuming that $h_{x_{i}}\rightarrow 0$. Obviously, now, this assumption must be relaxed, as $h_{x_{i}}\rightarrow \Delta$ where $\Delta$ contains a lot of objects that can be squared to zero. 
Let us continue our standard derivation of BRST quantisation. 
We introduce the ghosts $b_{A}$, $c^{\alpha}$, as well as a Lagrangian multiplier field $B_{A}$ so that we re-write our action as
\begin{widetext}
\begin{equation}
Z=\int [\mathcal{D}\phi \mathcal{D}B_{A}\mathcal{D}b_{A}\mathcal{D}b{a}\mathcal{D} c^{\alpha}]exp\{-S[\phi]+iB_{A}F^{A}[\phi]-b_{A}c^{\alpha}\delta_{\alpha}F^{A}[\phi]\}
\end{equation}
\end{widetext}
We implement gauge fixing, and include all ghosts, and therefore we obtain the gauge symmetry in the form of a fermionic global symmetry $\delta_{B}$ (BRST) together with the propagating ghosts which encode the local parts of the symmetry.
\begin{equation}
\begin{array}{c}
\delta_{B}\phi=-ic^{\alpha}\delta_{\alpha}\phi\\
\\
\delta_{B}B_{A}=0\\
\\
\delta_{B}b_{A}=B_{A}\\
\\
\delta_{B}c^{\alpha}=\frac{i}{2}f_{\beta\gamma}^{\alpha}c^{\beta}c^{\gamma}\\
\\
\end{array}
\end{equation}
which closes hence $\delta_{B}^{2}=0$ and therefore
\begin{equation}
iB_{A}F^{A}[\phi]-b_{A}c^{\alpha}\delta_{\alpha}F^{A}[\phi]=\delta_{B}(ib_{A}F^{A}[\phi])
\end{equation}
making $\delta_{B}$ a symmetry of the action. 
But not so fast! Our ghost fields are there to give us the linear corrections induced by our gauge transformations. We re-expressed the Jacobian to introduce them, and completed it in a fully supersymmetric form. Those ghosts are bound to having a unique zero, as expressed by standard calculus based on fundamental point-like structures. This is not the case when we replace our zeros with $\Delta$. Therefore, our ghosts can propagate not only in one single way to preserve the best local linear approximation to the gauge trajectory after gauge transformation, but in fact they can follow any possible path allowed by the replacement of our one single number $h$ of the derivative, with the object $\Delta$. If we patch our space (or our gauge space) with pieces of $\Delta$ tiles, our Jacobian will allow for basically a lot more transformations. Each partial derivative in our Jacobian will be associated with a whole set of infinitesimals that will have the possibility to be zero when squared in various directions on the manifold. Therefore our ghost fields will have a far more complex structure than the simple fields we assumed them to be. 
To make this clearer, suppose a function is $f:\mathbb{R}^{n}\rightarrow \mathbb{R}^{m}$ and we have first order partial derivatives well defined. The Jacobian will be of the form 
\begin{equation}
\begin{pmatrix}
\frac{\partial f_{1}}{\partial x_{1}}&\frac{\partial f_{1}}{\partial x_{2}}&\frac{\partial f_{1}}{\partial x_{3}}&...&\frac{\partial f_{1}}{\partial x_{n}}\\
\\
\frac{\partial f_{2}}{\partial x_{1}}&\frac{\partial f_{2}}{\partial x_{2}}&\frac{\partial f_{2}}{\partial x_{3}}&...&\frac{\partial f_{2}}{\partial x_{n}}\\
\\
\frac{\partial f_{3}}{\partial x_{1}}&\frac{\partial f_{3}}{\partial x_{2}}&\frac{\partial f_{3}}{\partial x_{3}}&...&\frac{\partial f_{3}}{\partial x_{n}}\\
\\
...&...&...&...&\\
\\
\frac{\partial f_{m}}{\partial x_{1}}&\frac{\partial f_{m}}{\partial x_{2}}&\frac{\partial f_{m}}{\partial x_{3}}&...&\frac{\partial f_{m}}{\partial x_{n}}\\
\end{pmatrix}
\end{equation}
This looks very simple only if we assume that there is only one possible differential limit, namely zero. But instead our limit is the differential object $\Delta$ which has a whole interval of elements that have the same property like zero required in this context, namely that they are having zero squares $x^{2}=0$. 

In order to have an idea about what this means in synthetic geometry, we should probably remember the fact that if we abandon the rule of excluded middle, as is the case in synthetic geometry, the notion of linear dependence is not so strict anymore. In particular, the existence of a non-zero Jacobian, allowing for the construction of our ghost fields and the propagation of the gauge choices has an additional intermediate situation, which has been completely ignored in classical geometry. The Jacobian is clearly also a measure of linear dependence of the partial derivatives of first order included. A transformation that would imply a null Jacobian will basically reduce one of the dimensions and would make the volume element resulting after the transformation collapse along one or several dimensions. We generally assume that this doesn't happen. However, in synthetic geometry there is an alternative situation, when we have a set of vectors that are not linearly free but are not linearly dependent either. 
Therefore we have an interesting situation that leads us to think at how and why string theory requires extra dimensions. The obvious explanation which was the standard explanation up to now is that in order to harmonise the Lorentz symmetry and the gauge symmetries for a string in the process of quantisation, the specific requirement for extra-dimensions emerged quite naturally from the cancellation requirements of the anomalies that are identified when ghost fields are used. However, the ghost fields have been created in the form of the Jacobian determinants making assumptions about the inherent logic of the geometry and about the unambiguous logical distinction between linear dependence and linear independence. In synthetic geometry a "middle" situation is in fact permitted, which leads to the annihilation of the resulting anomaly not by means of additional dimensions of the underlying spacetime but by means of the modification of the logic behind the geometry used. Instead of classical logic, intuitionistic logic allows for an intermediate situation in which the Jacobian transformation leading to the ghost fields systematically reduces the dimensions whenever a need for their increase appears. The reason for this reduction appears from the fact that the idea of "zero" is interpreted differently, not as a single entity, but as a region or equivalence class which may have structure and various topological and geometrical properties itself. 

Indeed, in order to complete our construction we had to make use of ghost fields that behave like Grassmann numbers, which do have this specific property, but that was associated to the fact that we expected some form of fermion degrees of freedom and hence were automatically thinking at supersymmetry. Of course, BRST is based on some form of supersymmetric completion of the field structure, but the actual origin and cause for this was not clearly understood. The role of the fermionic ghosts was there to actually close the algebra in the most general case adapted to the bosonic string. Of course, more advanced prescriptions are known in the case of the Batalin-Vilkoviski or field-anti-field prescriptions, but that is of little importance now. From a synthetic point of view, the requirement for a supersymmetric structure during the BRST quantisation seems to be related to the fact that the limit taken when studying an extended object (the string) was taken in the same way as when a point-like object was studied, and that limit simply made no sense for a string. 
Let us continue by defining a Noether current for the BRST symmetry, say $j_{B}$ and a corresponding $BRST$ charge, namely our $Q_{B}$ such that $\delta_{B}=i\{Q_{B},\cdot\}_{Poisson}$. We can see $Q_{B}$ now as a Hermitian quantum operator (given the integration measure remains well defined) and we obtain $Q_{B}^{2}=0$. 
Let us be even more specific and single out the types of transformations we need in our action functional. We have three independent types of transformations: two reparametrisation symmetries and the Weyl scaling, and therefore we will consider the independent components for the gauge transformations as 
\begin{equation}
\begin{array}{ccc}
h_{++}(\sigma), & h_{--}(\sigma), & h_{+-}(\sigma)
\end{array}
\end{equation}
The usual gauge choice made to construct a gauge slice that takes into account a particular choice for each of the three functions above would be
\begin{equation}
h_{\alpha \beta}=e^{\phi}\eta_{\alpha\beta}
\end{equation}
leading to $0=h_{++}=h_{--}$ in light cone coordinates. Given world sheet reparametrisations we have
\begin{equation}
\begin{array}{c}
\sigma^{+}\rightarrow \sigma^{+}+\xi^{+}\\
\sigma^{-}\rightarrow \sigma^{-}+\xi^{-}\\
\end{array}
\end{equation}
we have the gauge conditions
\begin{equation}
\begin{array}{c}
\delta h_{++}=2\nabla_{+}\xi_{+}\\
\\
\delta h_{--}=2\nabla_{-}\xi_{-}\\
\end{array}
\end{equation}
Basically we have the formula for the transformation of the metric tensor under infinitesimal coordinate reparametrisations
\begin{equation}
\delta h_{\alpha\beta}=\nabla_{\alpha}\xi_{\beta}+\nabla_{\beta}\xi_{\alpha}
\end{equation}
where we have the covariant derivative based on the Christoffel connection as 
\begin{equation}
\nabla_{\alpha}\xi_{\beta}=\partial_{\alpha}\xi_{\beta}-\Gamma^{\gamma}_{\alpha\beta}\xi_{\gamma}
\end{equation}
We can write this using the group of the reparametrisations of the worldsheet $G$ with the measure of integration $Dg$ over the group manifold and making the notation $h^{g}$ for the transformed metric by the reparametrisation $g$. Then the standard gauge fixing procedure implies introducing a unity in the form of the path integral that formalises the gauge fixing required
\begin{equation}
1=\int D g(\sigma)\delta(h_{++}^{g})\delta(h_{--}^{g})det(\frac{\delta h_{++}^{g}}{\delta g})\cdot det(\frac{\delta h_{--}^{g}}{\delta g})
\end{equation}
and we put it into the partition function as
\begin{widetext}
\begin{equation}
Z=\int D g(\sigma)\int D h(\sigma) D X(\sigma)e^{-S[h,X]}\delta(h_{++}^{g})\delta(h_{--}^{g})det(\frac{\delta h_{++}^{g}}{\delta g})\cdot det(\frac{\delta h_{--}^{g}}{\delta g})
\end{equation}
\end{widetext}
With a change of variables $h'=h^{g}$ we see that the action only depends on $h$ and $g$ in a combination defined by $h^{g}$ and hence by $h'$ and the integral over the group measure contributes with an infinite multiplicative factor only. 
The determinants are usually written in the following way, as integrals over anti-commuting ghosts $c^{-}$ and antighosts $b_{--}$
\begin{equation}
det(\frac{\delta h'_{++}}{\delta g})=\int D c^{-}(\sigma)Db_{--}(\sigma)exp\{-\frac{1}{\pi}\int d^{2}\sigma c^{-}\nabla_{+}b_{--}\}
\end{equation}
and the other determinant is an integral over the ghost $c^{+}$ and the anti-ghost $b_{++}$

\begin{equation}
det(\frac{\delta h'_{--}}{\delta g})=\int D c^{+}(\sigma)Db_{++}(\sigma)exp\{-\frac{1}{\pi}\int d^{2}\sigma c^{+}\nabla_{-}b_{++}\}
\end{equation}
We use the delta function in the partition function to solve for $h$ in terms of the conformal factor $\phi$ and write everything briefly 
\begin{equation}
Z=\int D\phi(\sigma)\int DX(\sigma) Dc(\sigma)D b(\sigma)e^{-S(X,b,c)}
\end{equation}
The decoupling of $\phi$ is linked to the cancelling of the conformal anomaly in the Virasoro algebra, which indicates an incompatibility between the three symmetries involved in the quantum description. With this mindset, the cancellation of this anomaly is only possible (for the bosonic string) in 26 background spacetime dimensions. However, we can see clearly that the determinants here are Jacobians that give us linear approximations for our transformations and rely on the existence of a single differential limit (zero). That is not true for extended objects. Therefore our construction of the determinants as path integrals over ghost fields as written above is insufficient. The situation is of course complicated by the fact that we do not have first order derivatives but instead first order functional dependences, and hence functional derivatives. But this complication is only minor. Instead of having a infinitesmial $\Delta$ defined in terms of a null squared interval, we have it represented in terms of functional forms. Given a manifold $M$ representing smooth functions $\rho$ and the functional $F$, $F:M\rightarrow R$ we have 
\begin{equation}
\int\frac{\delta F}{\delta \rho}(x)\phi(x)dx=\lim_{h\rightarrow 0}\frac{F[\rho+\epsilon \phi]-F[\rho]}{h}
\end{equation}
where $\phi$ is an arbitrary function and $h\cdot \phi$ is the small variation of $\rho$. 
Our differential is modified from representing an interval $\Delta$ where the condition $h^{2}=0$ is verified, to a region of our functional space where the small variation of $\rho$ namely $h\cdot \phi$  has that property. Instead of having a single such limit, we expand this limit into a domain that is considered to be minimal and fundamental. This amounts to the same conclusion as for the traditional Jacobian: we don't have only one possible way in which the limit $h\rightarrow 0$ can be taken, but instead we have $h\rightarrow \Delta$. This gives us more possibilities to locally eliminate the conformal anomaly, without the need for additional dimensions. Instead of considering only one way in which the limit can be taken, in particular only one number that is the infinitesimal limit (say zero) we can take a whole domain of possible deformations of the ghost fields within the domain where the square of our small parameter is null. 
To make this clearer, if we consider a functional $F[f]$ written as
\begin{equation}
F[f]=\int_{a}^{b}L(x,f(x),f'(x))dx
\end{equation}
in the standard way we consider $f'(x)=\frac{df}{dx}$, however we have seen that this description must be made valid for all quantities that can be squared to zero in the interval $\Delta$ that we use in synthetic geometry. Generalising, if $f$ is modified by the addition of a function $\delta f$ we obtain 
\begin{equation}
L(x,f+\delta f, f'+\delta f')
\end{equation}
and if we expand this in powers of our small quantity $\delta f$ the variation of the functional is well known 
\begin{widetext}
\begin{equation}
\begin{array}{c}
\delta F[f]=\int_{a}^{b}(\frac{\partial L}{\partial f}\delta f(x)+\frac{\partial L}{\partial f'}\frac{d}{dx}\delta f(x))dx=\\
\\
= \int _{a}^{b}(\frac{\partial L}{\partial f}-\frac{d}{dx}\frac{\partial L}{\partial f'})\delta f(x)dx+\frac{\partial{L}}{\partial f'}(b)\delta f(b)-\frac{\partial L}{\partial f'}(a)\delta f(a)
\end{array}
\end{equation}
\end{widetext}
which we recognise as the start of the Lagrangian mechanics for field theory. However, this entire construction relies on the fact that again, even in this space, the differential object $\delta f$ is a single deformation and its limit can be defined as going (in some integral sense) to a single zero. This implies there is only one object that is squared to become zero, and that object is zero. 
We need the ghosts as, while we have the anti-commuting relations that they obey
\begin{equation}
\begin{array}{c}
\{c_{m}, b_{m}\}=\delta_{m+n}\\
\\
\{c_{m},c_{n}\}=\{b_{m},b_{n}\}=0\\
\end{array}
\end{equation}
as fermionic fields, they have the property that $b^{2}=c^{2}=0$ which can be seen from the anti-commutation relations namely 
\begin{equation}
\{c_{m},c_{m}\}=c_{m}\cdot c_{m}+c_{m}\cdot c_{m}=2c_{m}^{2}=0
\end{equation}
This requirement appears to be a remnant from the fact that we designed our construction in an implicit sense like a synthetic geometric object, but we didn't consider that property in all instances, particularly not so when we performed the path integral or designed our theory in the usual Lagrangian way, by introducing Euler-Lagrange equations. In all these instances, some information has been lost regarding the way in which the infinitesimal limit in the construction of the first order derivatives has been taken. This is the actual reason why there exists a conformal anomaly in 4 dimensions, and probably not that the background space must necessarily be higher dimensional (which it could be, I don't know, but certainly the argument in favour of that will be much weaker after this material is understood). By introducing the fundamental string while keeping our differential calculus point-based, we only did half of the work, and obviously encountered inconsistencies. Let us see what can be done for the decoupling of the conformal factor $\phi$. Let us consider our Riemannian geometry for the case of the two dimensional worldsheet. The conformal form is 
\begin{equation}
h_{\alpha\beta}=e^{\phi}\eta_{\alpha\beta}
\end{equation}
We keep an Euclidean formulation, the rotation to Minkowski being relatively easy
\begin{equation}
ds^{2}=e^{\phi}(d\sigma^{2}+d\tau^{2})
\end{equation}
and we write in terms of complex coordinates $z=\tau+i\sigma$ and $\bar{z}=\tau-i\sigma$. We define
\begin{equation}
\begin{array}{c}
t_{\pm}=t_{0}\pm i t_{1}\\
\\
\partial_{\pm}=\frac{1}{2}(\frac{\partial}{\partial \tau} \mp i\frac{\partial}{\partial \sigma})\\
\end{array}
\end{equation}
with 
\begin{equation}
\begin{array}{c}
h_{++}=h_{--}=0\\
\\
h_{+-}=h_{-+}=\frac{1}{2}e^{\phi}\\
\end{array}
\end{equation}
and 
\begin{equation}
ds^{2}=e^{\phi}dz\cdot d\bar{z}
\end{equation}
we can use the conformal factor to raise and lower the indices 
\begin{equation}
\begin{array}{c}
t_{+}=\frac{1}{2}e^{\phi}t^{-}\\
\\
t_{-}=\frac{1}{2}e^{\phi}t^{+}\\
\end{array}
\end{equation}
In these complex coordinates a change $z\rightarrow z'=f(z)$ with $f$ a holomorphic function of $z$ will preserve the conformally flat form of the metric and given $\rho=e^{\phi}$ we have
\begin{equation}
\rho\rightarrow \rho'=\lvert \frac{dz'}{dz}\rvert^{-2}\rho
\end{equation}
and in general with $n_{u}$ upper and $n_{l}$ lower holomorphic indices and $\bar{n}_{u}$ upper and $\bar{n}_{l}$ lower antiholomorphic indices we have
\begin{equation}
t\rightarrow t'=(\frac{dz'}{dz})^{n_{u}-n_{l}}(\frac{d\bar{z'}}{d\bar{z}})^{\bar{n}_{u}-\bar{n}_{l}}t
\end{equation}
where the exponents are the holomorphic and antiholomorphic conformal dimensions of the tensor $t$. 
Using the standard covariant derivatives for tensors and the Christoffel connection 
\begin{equation}
\Gamma^{\gamma}_{\alpha\beta}=\frac{1}{2}h^{\gamma\delta}(\partial_{\alpha}h_{\beta\delta}+\partial_{\beta}h_{\alpha}{\delta}-\partial_{\delta}h_{\alpha\beta})
\end{equation}
and the Riemann curvature tensor
\begin{equation}
R^{\gamma}_{\alpha\beta\rho}=\partial_{\rho}\Gamma^{\gamma}_{\alpha\beta}+\Gamma^{\epsilon}_{\alpha\beta}\Gamma^{\gamma}_{\rho\epsilon}-(\beta\leftrightarrow \rho)
\end{equation}
we have only two non-zero Christoffel connection terms for the conformally flat metric
\begin{equation}
\begin{array}{c}
\Gamma^{+}_{++}=\partial_{+}\phi\\
\\
\Gamma^{-}_{--}=\partial_{-}\phi\\
\\
\end{array}
\end{equation}
we can then write the action for a general worldsheet metric and use the fields $c^{+}$ and $c^{-}$ as components of a vector field $c^{\alpha}$ while the anti-ghost fields behave like $b_{\alpha\beta}$
\begin{equation}
S_{ghost}=-\frac{i}{2\pi}d^{2}\sigma\sqrt{h}h^{\alpha\beta}c^{\gamma}\nabla_{\alpha}b_{\beta\gamma}
\end{equation}
We notice that we introduced the ghost fields to detect where the anomaly comes from, in a sense, to regularise our action with respect to potential quantum (conformal) anomalies. Indeed we found them, because we introduced fermionic ghost fields, which are objects that square to zero, by means of their anti-commutation laws, at the level of their modes, but this has been done in a limited way. However, not all of the structure emerging from expanding our zero limit to our synthetic geometric $\Delta$ interval has been taken into account. It is therefore not terribly surprising that when one adds worldsheet supersymmetry and extends the fermionic fields the number of dimensions required by criticality diminishes from 26 to 10. The reduction however doesn't quite follow what happens in Nature. First, if we have an $N=1$ supersymmetry on the worldsheet, we do obtain D=10 spacetime target dimensions for criticality, but when we extend this to $N=2$ worldsheet supersymmetry what we obtain is the critical dimension D=2. Worse, $N=4$ worldsheet supersymmetry leads to a negative critical dimension. In any case, this type of supersymmetry extension doesn't seem to be the solution, and the alternative worked upon until now, namely the compactification of extra dimensions doesn't seem to be very predictive either. It seems as if something is missing in the quantisation procedure via BRST, but it is quite insightful to note that more supersymmetry (and additional fermionic fields that amount to additional quantities that square to zero) seem to impact the critical dimension. It is interesting to observe that the null square property appears in all these approaches: fermion modes square to zero due to the anticommutation relations, the BRST charge squares to zero because it is a closed boundary operator, and compactification does exactly the same thing, it introduces closed dimensions, which contribute each with closed boundary operators. There are however more interesting objects that produce such square zero results, and they emerge from the existence of infinitesimals that allow us to define the functional Jacobian as a generalised local linearisation. In differential synthetic geometry the infinitesimal is defined in terms of a line which is isomorphic to the smooth line $R$. Therefore it is easy to define a concept of derivative as it would be just the map that would send the extended infinitesimal that we defined as a line to a number equivalent to the value of the tangent in that point. Unfortunately strings as fundamental objects have much more structure and must be quantised. Therefore we cannot simply define the Jacobian in the same way we do it usually, also in synthetic differential geometry. The fact that our fundamental infinitesimal has quite significant structure by itself however requires modifications of the ghost structure in order to accommodate it. 
First when we started using synthetic geometry first order differentials, we lost some of the properties and even the logical structure of usual geometry. Even if we do not strictly adhere to the synthetic geometry approach to associate to every differential object a line object (because strings can vibrate for example) we cannot employ a certain logical construction, like the law of the excluded middle. In fact the geometrical construction behind the existence of non-trivial differential objects is the theory of toposes. Mathematically this is a very interesting subject, but only some very specific aspects of it will be used directly here. In any case, in bosonic string quantisation, the conformal anomaly being made manifest by ghost fields doesn't necessary indicate that the number of dimensions we are working with is wrong. It can also indicate that the way we are considering the extended objects and their gauge structure is wrong. In fact, both may be wrong. There is hardly a way to figure that out without experimental evidence, but there is reason to believe that there is an additional problem related to how we describe the quantisation prescription of extended objects. 
There is an interesting way of looking at synthetic geometry. 
\section{dimensionality and synthetic geometry}
We have seen in the previous section that the constraint that demands a certain dimension for the background spacetime in string theory can be relaxed by applying synthetic arguments to the differential geometric problem of quantisation of the string by means of the method of BRST cohomology. String theory proper was the first theory that demanded a certain dimension of the background space for the consistent quantisation of a theory with both Lorentz and internal gauge symmetry as well as conformal symmetry on the worldsheet. This constraint that was previously unknown has been altered by the modification of the internal logic and the structure of the infinitesimal due to synthetic geometry. Therefore, historically, we went from a situation in which the dimensions of spacetime were imposed on us from strictly empirical observations (we can still only "see" 3 spatial and 1 time dimensions), through a situation in which the extension of the fundamental objects from point particles to strings required a fixed higher spacetime dimension, to another situation in which the structure of the algebra of infinitesimals and of synthetic differential geometry was capable of relaxing the conditions within the BRST quantisation of string theory that led to the requirement of a certain dimension for the background of string theory. 
Thinking about this, we have to remember that string theory depends in its phenomenology on the existence and very specific compactification of the extra-dimensions. Otherwise stated, there is a lot of information in the extra dimensions that is suggesting that many of the properties of the standard model that have no clear explanation within the standard model can be understood by the dynamics of, for example, flux compactification. This is not new, from the Higgs mechanism appearing from tachyon decays of intersecting D-branes, to the hopeful explanation of lepton generations within string phenomenology, in all cases extra dimensions and their compactifications play a significant role in the understanding of possible standard model effects. If we are to say that the choice of a certain dimension is not fixed in synthetic geometry, then such determinations (granted, incomplete, as string theory itself cannot yet pinpoint the exact compactification that would lead to the standard model, nor does it know whether such a compactification would be special in any sense or not, see for example the literature on string multiverses for an introduction on this problem) should better exist in a synthetic description of the string. Indeed they do. However, this article would expand considerably if I were to derive the results here. Sufficient to say that the work of Vinogradov and his collaborators [11], [12] results in a categorification of geometry which then can be implemented for the synthetic geometry described here. I strongly suspect that the information related to compactification can be transferred to the categorical properties of the differential geometry or the higher differential geometry involved in the construction of synthetic string theory. Moreover, the methods used there may provide even stronger constraints that would lead to narrowing down the standard model as more necessary and less arbitrary when thinking in terms of the multiverse. Such synthetic / categorical constraints may originate from the map structure of the Vinogradov category of differential equations. It is very interesting to explore what constraints on possible synthetic string constructions would emerge from that. 
Another interesting remark is that the information stored in a traditional string theory through higher dimensions and compactification may equally well (or even better) be stored via a synthetic string theory through the cohomological properties emerging from the algebraic differential structure introduced. Therefore, such a synthetic theory would give us a better way in which spacetime dimensionality would emerge, namely from the homological algebraic construction at the level of the infinitesimal completion in synthetic differential geometry. 

\section{Practical future applications}
In a more speculative note, the idea that including intuitionistic logic, abandoning the law of excluded middle and introducing Toposes as a foundational structure in geometry has led to a massive reduction of the dimensions in which string theory can be formulated in a consistent way leads us to think about the possibility of employing the same mechanisms in other situations where the reduction of dimensionality is needed and welcomed. One such example is the dimension of the feature space in neural networks. There too, the dimension of the space must be reduced to more practical values without loosing the desired information. The procedures of dimensional reduction involve several methods which always do in fact eliminate some information. The desired result is that the information that has been eliminated is not the one that is relevant for the analysis being performed. 
However, if the notion of weak inversion and the intuitionistic logic of Toposes can be used to find correlations that amount to specific dimensions being reduced, then the information would be maintained in a lower dimension by means of intuitionistic logical connections on the Topos. As such, this type of representation could reduce the otherwise large dimensions involved in feature characterisations, often reaching values of thousands or more.
\section{conclusion}
In this article I presented a series of motivations for asking whether modified groups structures and a Synthetic geometric interpretation can be used to explain dualities and to obtain a string theoretical framework in lower dimensions. I also showed that the origin of dualities in string and gauge theory can be related to a quantum interpretation of the structure of theories themselves, leading to a situation in which dualities are equivalent with apparently distinct but formally inseparable theories in a specific theory space. This is seen as a quantum structure on top of our theory space that results in new dualities linking apparently different theories. 
\section{data availability}
This manuscript has no associated data

\end{document}